\documentclass[preprint]{ptephy_om}

\preprintnumber{KYUSHU-HET-371} 

\usepackage{graphics}

\usepackage{amsmath} 
\usepackage{amsthm} 
\usepackage{url} 
\usepackage{stmaryrd}
\usepackage{tikz}
\usepackage{tikz-feynhand}
\usepackage{mathtools}
\usepackage{bm}
\usepackage{booktabs}
\usepackage{subcaption}
\usepackage{color}

\usepackage{pgfplots}
\pgfplotsset{compat=1.18}
\usepgfplotslibrary{fillbetween}

\DeclareMathOperator{\imag}{Im}

\DeclareMathOperator{\Tr}{Tr}

\numberwithin{equation}{section}

\allowdisplaybreaks

\begin{document}

\title{Quantum Tunneling from Perturbation Theory Revisited}


\author[1,2]{Hiroshi Suzuki}
\affil[1]{Department of Physics, Kyushu University, 744 Motooka, Nishi-ku,
Fukuoka 819-0395, Japan}
\affil[2]{Quantum and Spacetime Research Institute (QuaSR), Kyushu University,
744 Motooka, Nishi-ku, Fukuoka 819-0395, Japan}





\begin{abstract}%
In the late 1990s, Suzuki and Yasuta proposed a compact formula that extracts
the decay rate per unit volume of a false vacuum in the $D$-dimensional $O(N)$
$\lambda(\phi^2)^2$ theory with an unbounded potential from conventional
perturbative coefficients of the vacuum energy density, i.e.\ vacuum bubble
diagrams. The idea was to identify the imaginary part arising from the Borel
integral along the discontinuity of the Borel transform with that of the vacuum
energy density. While the formula works quite well for~$D=1$, i.e.\ quantum
mechanics, its validity remained unclear for~$D\geq2$ because only the first
three perturbative coefficients for~$D=2$ were available and the result showed
no sign of convergence. In the present paper, we reexamine this approach using
the first seven nontrivial perturbative coefficients (up to nine loops)
for~$D=2$ and~$N=1$ obtained by Serone, Spada, and~Villadoro. Introducing two
tunable parameters in the finite-order truncated Borel transform following
these authors, we find that the imaginary part converges as the perturbative
order increases, with the last few orders agreeing to within a few percent. In
the intermediate range of the coupling constant,
$3\lesssim\widetilde{g}\lesssim8$, this approach yields an imaginary part
rather close to the leading-order semiclassical approximation with the
one-loop determinant; it is $10$--$20\%$ larger than the semiclassical result
with the two-loop correction computed by Malatesta, Parisi, and~Rizzo. Our
result is consistent, within errors, with the continuum extrapolation of the
Pad\'e-resummed linked-cluster expansion in a quantum $\phi^4$ chain.
\end{abstract}

\subjectindex{B30,B34,B87}

\maketitle

\section{Introduction}
\label{sec:1}
The interplay between perturbation theory and nonperturbative phenomena is
quite intriguing. It was recognized early on that the convergence of the
perturbation series is intimately related to the instability of the vacuum
under the sign flip of the coupling constant~\cite{Dyson:1952tj}, and that
tunneling phenomena mediated by instantons obstruct the Borel summability of
the perturbation
series~\cite{tHooft:1977xjm,Bogomolny:1980ur,Zinn-Justin:1981qzi}. From a
modern perspective, such an interplay between perturbative and nonperturbative
physics has been organized under the notions of resurgence and
trans-series~\cite{Aniceto:2018bis}. In quantum mechanics, the relation between
the Borel resummation of the perturbative series, Stokes phenomena, and
nonperturbative contributions has been clarified in detail by the exact
Wentzel--Kramers--Brillouin (WKB) analysis combined with
resurgence~\cite{Sueishi:2020rug,Sueishi:2021xti,Bucciotti:2023trp,Misumi:2025ijd}, which has recently been extended to unstable systems with resonances and
metastable
states~\cite{Kamata:2023opn,Morikawa:2025grx,Morikawa:2025xjq,Kamata:2026mjh}.
In this framework, the imaginary part of the energy of a metastable state
arises from the lateral Borel resummation of the perturbative series across the
Stokes line. The approach of the present paper may be regarded as a numerical
implementation of this picture in field theory, where the exact WKB analysis is
not available: we construct the lateral Borel sum directly from a finite number
of perturbative coefficients. See also~Refs.~\cite{Ai:2019fri,Garbrecht:2024end}.

Let us take the $D$-dimensional $O(N)$ $\lambda(\phi^2)^2$ theory ($D<4$ for
super-renormalizability) whose Euclidean action is
\begin{equation}
   S=\int d^Dx\,\left[
   \frac{1}{2}\partial_\mu\phi\partial_\mu\phi
   +\frac{1}{2}m^2\phi^2
   -\frac{1}{4!}g(\phi^2)^2
   \right].
\label{eq:(1.1)}
\end{equation}
Here, we consider the case $m^2>0$ and~$g>0$ so that, in our convention, the
potential energy is unbounded from below. The classical vacuum~$\phi=0$ is
metastable and eventually decays via quantum tunneling. The decay width per
unit volume is given by~${\mit\Gamma}=-2\imag{\mathcal{E}}/\hbar$, where
$\mathcal{E}$ is the ground-state (i.e.\ vacuum) energy density defined by the
analytic continuation from~$g<0$.

Usually, such a tunneling phenomenon is regarded as nonperturbative because
for~$\widetilde{g}:=g/m^{4-D}\ll1$, the semiclassical approximation (the
bounce or instanton calculus~\cite{Coleman:1978ae})
gives~\cite{Brezin:1978,Malatesta:2017}
\begin{equation}
   \left[\imag\widetilde{\mathcal{E}}(\widetilde{g})\right]_{\text{bounce}}
   =-A_NC_{D,N}\left({S_0\over2\pi\widetilde{g}}\right)^{(D+N-1)/2}
   e^{-S_0/\widetilde{g}}.
\label{eq:(1.2)}
\end{equation}
Here and in what follows, we use dimensionless combinations,
\begin{equation}
   \widetilde{\mathcal{E}}:=\mathcal{E}/m^D,\qquad
   \widetilde{g}:=g/m^{4-D}.
\label{eq:(1.3)}
\end{equation}
In Eq.~\eqref{eq:(1.2)}, $S_0$ is the bounce action,
\begin{equation}
   S_0=\begin{cases}
   8&D=1,\\
   35.10269&D=2,\\
   113.38351&D=3,\\
   \end{cases}
\label{eq:(1.4)}
\end{equation}
$A_N=\pi^{N/2}/\Gamma(N/2)$ is the moduli volume of the bounce, and
\begin{equation}
   C_{D,N}
   =\begin{cases}
   2\sqrt{3}\cdot(2\sqrt{3})^{N-1}&D=1,\\
   0.43245\cdot(1.652)^{N-1}&D=2,\\
   10.189\cdot(1.6569)^{N-1}&D=3,\\
   \end{cases}
\label{eq:(1.5)}
\end{equation}
is the one-loop determinant arising from the integration of fluctuations around
the bounce solution.\footnote{%
Here, for~$D=2$ and~$N=1$, we quote the value obtained
from~Ref.~\cite{Malatesta:2017}, which differs from the one
of~Ref.~\cite{Brezin:1978}; the authors of~Ref.~\cite{Malatesta:2017} attribute
the discrepancy to a possible misprint in~Ref.~\cite{Brezin:1978}. The value
of~Ref.~\cite{Brezin:1978} would instead give $C_{2,1}=0.3503$. In this
connection, the mass counterterm in the scheme in~Ref.~\cite{Malatesta:2017}
is~$\delta m^2=-3gG_0(0)+6g^2K(0)$ (in the notation of that reference). Its
$O(g)$ part, $-3gG_0(0)$, coincides with the tadpole subtraction in the normal
ordering scheme, so the value of~$C_{2,1}$ quoted here
from~Ref.~\cite{Malatesta:2017} can be used as it is in the normal ordering
scheme. The two schemes differ only at~$O(g^2)$, through the sunset
diagram~$K(0)$; this difference is taken into account
in~Eq.~\eqref{eq:(3.5)}.} Since $\imag\widetilde{\mathcal{E}}$ behaves
as~$\sim e^{-S_0/\widetilde{g}}$, the power series expansion
of~$\imag\widetilde{\mathcal{E}}$ vanishes to all orders in~$\widetilde{g}$.
This would imply that the tunneling is invisible in perturbation theory.

However, the Borel summability of the
system~\eqref{eq:(1.1)}~\cite{Loeffel:1969rdm,Simon:1970mc,Graffi:1970erh,Eckmann:1975agu,Magnen:1977ha} indicates that the Borel summation of the perturbative
series with respect to~$g$, after the analytic continuation from~$g<0$ to~$g>0$
(note our convention), would yield the imaginary part. In the picture based on
Lefschetz thimbles~\cite{Witten:2010cx,Witten:2010zr,Fujimori:2018nvz},
for~$g<0$, only the unique real saddle point~$\phi=0$ contributes, implying
Borel summability~\cite{Serone:2017nmd}, whereas for~$g>0$ perturbative
contributions around infinitely many saddle points given by (multi-)bounce
solutions should be summed over. These two cases, however, should give the same
answer. In this way, in the system~\eqref{eq:(1.1)}, the conventional
perturbation theory around~$\phi=0$ and the semiclassical calculation provide
two facets of a single object.

Now, in the late 1990s, inspired by the success of the variational perturbation
theory in tunneling problems in quantum mechanics~\cite{Kleinert:1992tq,Karrlein:1993ej,Kleinert:1995ii}\footnote{%
For this type of perturbation theory with a converging behavior, see also
Refs.~\cite{Seznec:1979ev,LeGuillou:1983,Guida:1994zv,Guida:1995px,Kleinert:1995hc}.} and intuitively expecting the above complementarity of the
perturbative and nonperturbative pictures, Suzuki and Yasuta proposed a
compact formula that extracts the decay width of a false vacuum from
conventional perturbative coefficients of the vacuum energy density, in quantum
mechanics ($D=1$ and~$N=1$ in~Eq.~\eqref{eq:(1.1)})~\cite{Suzuki:1996rt}, and in
$D$-dimensional $O(N)$ cases ($D<4$)~\cite{Suzuki:1997hb}. See
also~Ref.~\cite{Yasuta:1997it}. The idea of~Refs.~\cite{Suzuki:1996rt,Suzuki:1997hb} was to apply the Borel resummation literally to this problem,
supplemented with the conformal mapping method for the Borel integration
variable~\cite{Loeffel:1976}, and to identify the imaginary part arising from
the Borel integral along the discontinuity of the Borel transform with that of
the vacuum energy density. While it was found that the formula works quite
well for~$D=1$ (this is reproduced in~Appendix~\ref{sec:A}), its validity
remained unclear for~$D\geq2$ because only the first three perturbative
coefficients for~$D=2$ were available and the result showed no sign of
convergence.

In the present paper, we reexamine this approach using the first seven
nontrivial perturbative coefficients for~$D=2$ and~$N=1$ obtained
in~Ref.~\cite{Serone:2018gjo}. Introducing two tunable parameters in the
finite-order truncated Borel transform following Ref.~\cite{Serone:2018gjo}
and choosing them in the region where the result is stationary, we obtain an
imaginary part that converges with respect to the perturbative order, with the
last few orders agreeing to within a few percent. In the intermediate range of
the coupling constant, $3\lesssim\widetilde{g}\lesssim8$, our approach yields
an imaginary part rather close to the leading-order semiclassical
approximation with the one-loop determinant; it is $10$--$20\%$ larger than the
semiclassical result with the two-loop correction computed
in~Ref.~\cite{Malatesta:2017}. If taken at face value, this result would
indicate that, in the intermediate coupling region, the higher-order
corrections in the semiclassical expansion are not small.

This paper is organized as follows. In~Section~\ref{sec:2}, we recapitulate the
formulation of~Ref.~\cite{Suzuki:1997hb}, generalized by two tunable parameters
$b$ and~$s$ in the finite-order truncated Borel
transform~\cite{Serone:2018gjo}. Our basic formula for the imaginary part from
the perturbative series is~Eq.~\eqref{eq:(2.12)}. In~Section~\ref{sec:3}, we
present results of the numerical experiment on the basis
of~Eq.~\eqref{eq:(2.12)}. We first confirm that the choice of the parameters
$b=1/2$ and~$s=0$ reproduces the results in~Ref.~\cite{Suzuki:1997hb} that show
no sign of convergence; see~Fig.~\ref{fig:1} in~Section~\ref{sec:3}. Then, we
basically follow the strategy of~Ref.~\cite{Serone:2018gjo}, i.e.\ an optimal
choice of~$b$ and~$s$. In the present paper, however, we fix $b$ by~Eq.~\eqref{eq:(2.15)}, which reproduces the weak coupling behavior of the imaginary part.
We tune the other parameter~$s$ that controls the strong coupling behavior by
requiring that it minimizes the quantity~\eqref{eq:(3.3)}, following the
criterion of~Ref.~\cite{Serone:2018gjo}. Then, as~Fig.~\ref{fig:4}
in~Section~\ref{sec:3} shows, at least in the intermediate range of the
coupling constant, $3\lesssim\widetilde{g}\lesssim8$, we observe a clear
tendency toward convergence. The error associated with the choice of the
parameter is estimated by~Eq.~\eqref{eq:(3.4)}. Section~\ref{sec:4} is devoted
to the conclusion. In~Appendix~\ref{sec:A}, we carry out an analogous numerical
experiment for a quantum mechanical case, $D=1$ and~$N=1$, for which an exact
result can be obtained, to illustrate the validity of the strategy in the main
text. In~Appendix~\ref{sec:B}, we attempt to compute the imaginary part
independently, by the continuum extrapolation of the Pad\'e-resummed
linked-cluster expansion in a quantum $\phi^4$ chain.

\section{Formulation}
\label{sec:2}
We set the conventional perturbative expansion of the vacuum energy density in
the system~\eqref{eq:(1.1)} as
\begin{equation}
   \widetilde{\mathcal{E}}(\widetilde{g})
   \sim\sum_{n=0}^\infty c_n\widetilde{g}^n.
\label{eq:(2.1)}
\end{equation}
Here, we have used $\sim$ rather than the equality sign~$=$ to emphasize that
this series is not convergent but at best asymptotic. Diagrammatically, the
coefficients~$c_n$ are given by vacuum bubble diagrams.

The series~\eqref{eq:(2.1)}, although divergent, contains useful physical
information. It may be extracted by introducing the Borel transform:
\begin{equation}
   B_b(z):=\sum_{n=0}^\infty\frac{c_n}{\Gamma(n+b+1)}z^n,
\label{eq:(2.2)}
\end{equation}
where $b>-1$ is a free parameter; later we use this freedom to improve the
convergence of the result. Then from the large-order behavior of the
perturbative coefficients,
\begin{equation}
   c_n\sim-\frac{A_NC_{D,N}}{\pi(2\pi)^{(D+N-1)/2}}
   \frac{\Gamma(n+(D+N-1)/2)}{S_0^n}
\label{eq:(2.3)}
\end{equation}
for~$n\gg1$ that can be deduced from the weak coupling behavior of the
imaginary part, Eq.~\eqref{eq:(1.2)},\footnote{%
Equation~\eqref{eq:(2.3)} can be obtained by substituting Eq.~\eqref{eq:(1.2)}
into~$c_n=\frac{1}{\pi}\int_0^\infty\frac{d\widetilde{g}'}{\widetilde{g}^{\prime(n+1)}}\,\imag\widetilde{\mathcal{E}}(\widetilde{g}')$, which is derived by
expanding the dispersion relation without subtraction,
$\widetilde{\mathcal{E}}(\widetilde{g})
=\frac{1}{\pi}\int_0^\infty d\widetilde{g}'\,
\frac{\imag\widetilde{\mathcal{E}}(\widetilde{g}')}
{\widetilde{g}'-\widetilde{g}}$ in~$g$. For~$n\gg1$, the integral is dominated
by~$\imag\widetilde{\mathcal{E}}(\widetilde{g}')$ for~$\widetilde{g}'\ll1$.
See~Ref.~\cite{LeGuillou:1990nq} and references therein.} we see that the
series~\eqref{eq:(2.2)} now possesses a finite radius of convergence. The
radius of convergence is given by the singularity nearest to the origin
at~$z=S_0$:\footnote{%
In~Refs.~\cite{Suzuki:1996rt,Suzuki:1997hb}, the parameter~$b$ is fixed
as~$b=(D+N)/2-1$. This choice leads to~$\gamma=1/2$ and makes the branch point
at~$S_0$ an integrable singularity.}
\begin{equation}
   B_b(z)=-{A_NC_{D,N}\over\pi(2\pi)^{(D+N-1)/2}}\Gamma(\gamma)
   \left(1-\frac{z}{S_0}\right)^{-\gamma}+\cdots,\qquad
   \gamma:=\frac{D+N-1}{2}-b.
\label{eq:(2.4)}
\end{equation}

From the Borel transform~\eqref{eq:(2.2)}, the Borel sum of the divergent
series~\eqref{eq:(2.1)} is defined by
\begin{equation}
   \widetilde{\mathcal{E}}(\widetilde{g})
   :=\frac{1}{\widetilde{g}^{b+1}}
   \int_0^\infty dz\,e^{-z/\widetilde{g}}
   z^b B_b(z+i\varepsilon).
\label{eq:(2.5)}
\end{equation}
If the series is of a finite number of terms,
$\widetilde{\mathcal{E}}(\widetilde{g})=\sum_{n=0}^Pc_n\widetilde{g}^n$, one can
confirm that the above procedures, Eqs.~\eqref{eq:(2.2)} and~\eqref{eq:(2.5)},
give back the original series. In favorable cases, the Borel
sum~\eqref{eq:(2.5)} yields a finite result even if the series~\eqref{eq:(2.1)}
is divergent. However, in the present case of nonsignalternating
coefficients~\eqref{eq:(2.3)}, the Borel transform~$B_b(z)$ develops
singularities on the \emph{positive\/} real axis as in~Eq.~\eqref{eq:(2.4)},
and we have to deform the integration contour to define the integral.
In~Eq.~\eqref{eq:(2.5)}, we have deformed the integration path
as~$z\to z+i\epsilon$ ($\epsilon>0$ is infinitesimal) so that it passes above
the singularities on the positive real axis. This deformation of the Borel
integral produces the imaginary part that we identify with the physical decay
width of the false vacuum. We will find that this choice of the deformed
contour gives a \emph{negative\/} imaginary part, i.e.\ a positive decay width,
and thus this choice is in accord with the physical situation, i.e.\ the decay
of the false vacuum.

In order to carry out the integration~\eqref{eq:(2.5)}, however, we have to
know the Borel transform~$B_b(z)$~\eqref{eq:(2.2)} for $0\leq z<\infty$, i.e.\
even outside the circle of convergence of the series~\eqref{eq:(2.2)}. One has
to analytically continue the function~$B_b(z)$~\eqref{eq:(2.2)} beyond the
circle of convergence~$|z|=S_0$ but this would be impossible without knowing
infinitely many terms of~Eq.~\eqref{eq:(2.2)}.

This difficulty of the analytic continuation may be avoided by introducing a
conformal mapping for the integration variable~\cite{Loeffel:1976},
\begin{equation}
   z=4S_0\frac{\lambda}{(1+\lambda)^2},\qquad
   \lambda=\frac{1-\sqrt{1-z/S_0}}{1+\sqrt{1-z/S_0}}.
\label{eq:(2.6)}
\end{equation}
This maps the complex $z$-plane with the cut along~$(S_0,\infty)$ onto the
disk~$|\lambda|\leq1$. The cut corresponds to the circle~$|\lambda|=1$; the
point $z=S_0$ is mapped to~$\lambda=+1$ and~$z=\infty$ to~$\lambda=-1$. In
this way, the series~\eqref{eq:(2.2)} in terms of~$\lambda$ converges within
the disk~$|\lambda|<1$ within which the integration contour
in~Eq.~\eqref{eq:(2.5)} is now confined.\footnote{%
\label{footnote:5}
Here, we have to assume that there is no singularity in the cut z-plane, i.e.\
outside $[S_0,\infty)$. Although we are not aware of a proof of this, the
success of the Borel summation of correlation functions
in~Ref.~\cite{Serone:2018gjo} by the same conformal mapping strongly indicates
that this is the case.}

Also, by using the variable~$\lambda$, we rewrite Eq.~\eqref{eq:(2.5)} as
\begin{equation}
   \widetilde{\mathcal{E}}(\widetilde{g})
   =\frac{1}{\widetilde{g}^{b+1}}
   \int_0^\infty dz\,e^{-z/\widetilde{g}}
   z^b(1+\lambda)^{-2s}B_{b,s}(z),
\label{eq:(2.7)}
\end{equation}
where $s$ is another free parameter~\cite{Serone:2018gjo} and
\begin{equation}
   B_{b,s}(z):=(1+\lambda)^{2s}B_b(z).
\label{eq:(2.8)}
\end{equation}
By construction, Eq.~\eqref{eq:(2.7)} as it stands does not depend on the
parameter~$s$. However, if we expand $B_{b,s}(z)$~\eqref{eq:(2.8)} with respect
to~$\lambda$,
\begin{equation}
   B_{b,s}(z)=\sum_{\ell=0}^\infty d_\ell\lambda^\ell
\label{eq:(2.9)}
\end{equation}
and \emph{truncate the series at a finite order}, say~$P$, then
Eq.~\eqref{eq:(2.7)} depends on~$s$. We will use this freedom in the
resummation also to improve the convergence of the
result~\cite{Serone:2018gjo}. It turns out that the introduction of the
parameter~$s$ improves the situation considerably.

The coefficients~$d_\ell$ in~Eq.~\eqref{eq:(2.9)} are obtained from the original
coefficients~$c_n$ in~Eq.~\eqref{eq:(2.1)} by
\begin{equation}
   d_\ell:=\sum_{n=0}^\ell(4S_0)^n
   \begin{pmatrix}
   2s-2n\\
   \ell-n\\
   \end{pmatrix}
   \frac{c_n}{\Gamma(n+b+1)},
\label{eq:(2.10)}
\end{equation}
where
\begin{equation}
   \begin{pmatrix}
   \alpha\\
   m\\
   \end{pmatrix}
   :=\frac{1}{m!}\prod_{j=0}^{m-1}(\alpha-j)
\label{eq:(2.11)}
\end{equation}
is the generalized binomial coefficient.

Finally, after the change of integration variable from~$z$ to~$\lambda$, we
parametrize the integration contour on the $\lambda$-plane, the upper half
circle, as~$\lambda=e^{i\theta}$, where~$0\leq\theta\leq\pi$.\footnote{%
The other segment of the integration path, $0\leq\lambda<1$, can be ignored
because it does not produce any imaginary part.} Then, taking the imaginary
part of~Eq.~\eqref{eq:(2.7)} and truncating the series at the $P$th order, we
have our final formula,\footnote{%
The corresponding formula in~Ref.~\cite{Suzuki:1997hb} is given by
setting~$b=(D+N)/2-1$ and~$s=0$.}
\begin{align}
   &\left[\imag\widetilde{\mathcal{E}}(\widetilde{g})\right]_P(b,s)
\notag\\
   &=\left(\frac{S_0}{\widetilde{g}}\right)^{b+1}
   2^{-2s}
   \int_0^\pi d\theta\,
   \exp\left(-\frac{S_0}{\widetilde{g}}\frac{1}{\cos^2\theta/2}\right)
   \frac{\sin\theta/2}{\cos^{2b+3+2s}\theta/2}
   \sum_{\ell=0}^Pd_\ell\sin(\ell-s)\theta,
\label{eq:(2.12)}
\end{align}
where the perturbative coefficients~$d_\ell$ are given by~Eq.~\eqref{eq:(2.10)}.
Here, we have indicated that this formula depends on two parameters, $b$
and~$s$, which may be employed to improve the convergence with respect to the
approximation order~$P$.

Now, setting~$u=\tan(\theta/2)$ in~Eq.~\eqref{eq:(2.12)}, we have
\begin{align}
   &\left[\imag\widetilde{\mathcal{E}}(\widetilde{g})\right]_P(b,s)
\notag\\
   &=\left(\frac{S_0}{\widetilde{g}}\right)^{b+1}
   2^{-2s+1}
   \int_0^\infty du\,
   \exp\left[-\frac{S_0}{\widetilde{g}}(1+u^2)\right]
   (1+u^2)^{b+s}u
   \sum_{\ell=0}^Pd_\ell\sin(\ell-s)\theta,
\label{eq:(2.13)}
\end{align}
where $\theta=2\arctan u$. For a fixed order~$P$, we see that the saddle point
of the integrand is given by~$u^2\sim\widetilde{g}/S_0$ in either case,
$\widetilde{g}\to0$ or~$\widetilde{g}\to\infty$. At the saddle point, we may
set, retaining only the dependence on~$\widetilde{g}$,
$\sin(\ell-s)\theta\sim\widetilde{g}^{1/2}$ for the former case and
$\sin(\ell-s)\theta\sim\text{const.}$ for the latter case. From these
considerations, we see that the asymptotic behavior is given
by\footnote{%
It is necessary to take into account the Gaussian integration around the
saddle point, which produces a factor~$\widetilde{g}^{1/2}$.}
\begin{equation}
   \left[\imag\widetilde{\mathcal{E}}(\widetilde{g})\right]_P(b,s)
   \sim\begin{cases}
   \widetilde{g}^{-b+1/2}\exp(-S_0/\widetilde{g})&\widetilde{g}\to0,\\
   \widetilde{g}^s&\widetilde{g}\to\infty.\\
   \end{cases}
\label{eq:(2.14)}
\end{equation}
From this, we realize that the parameter~$b$ controls the behavior of the
approximation in the weak coupling limit~$\widetilde{g}\to0$, whereas $s$
controls the behavior in the strong coupling limit. In this paper, we choose
\begin{equation}
   b=\frac{D+N}{2},
\label{eq:(2.15)}
\end{equation}
so that the power of~$\widetilde{g}$ in~Eq.~\eqref{eq:(2.14)} and that
in~Eq.~\eqref{eq:(1.2)} coincide. In the numerical experiment in the next
section, we see that this choice works quite well.

Let us clarify the nonperturbative information assumed in the present
approach. The conformal mapping~\eqref{eq:(2.6)} uses the position~$S_0$ of the
singularity of~$B_b(z)$ nearest to the origin, and the choice~\eqref{eq:(2.15)}
matches the power of~$\widetilde{g}$ in Eq.~\eqref{eq:(1.2)}. The
normalization~$A_NC_{D,N}$ and all the corrections in~$\widetilde{g}$ are, on
the other hand, not assumed; they are obtained from the perturbative
coefficients. In fact, both $S_0$ and~$C_{2,1}$ can already be read off from the
coefficients~\eqref{eq:(3.2)} themselves: $(n+1)c_n/c_{n+1}=34.4$--$35.3$
for~$3\leq n\leq7$, to be compared with~$S_0=35.10$, and
$-2\pi^2c_nS_0^n/n!=0.429$--$0.443$ for~$3\leq n\leq8$, to be compared
with~$C_{2,1}=0.43245$ (or $0.3503$ of~Ref.~\cite{Brezin:1978});
see~Eq.~\eqref{eq:(2.3)}. Other singularities of the Borel transform, such as
those at~$z=kS_0$ ($k=2$, $3$, \dots) associated with multibounce
configurations, lie on the same cut and are mapped onto the unit
circle~$|\lambda|=1$. Since the theory is super-renormalizable and massive,
there are no renormalon singularities, and the Borel summability
for~$g<0$~\cite{Loeffel:1969rdm,Simon:1970mc,Graffi:1970erh,Eckmann:1975agu,Magnen:1977ha} excludes singularities on the negative real axis. The absence of
singularities elsewhere in the cut plane is an assumption
(see~footnote~\ref{footnote:5}).

For the parameter~$s$, on the other hand, we leave it as a tunable parameter
and use this freedom to make the results insensitive to the choice of~$s$ and
to the difference of the approximation order~$P$. In practice, we use $s$ that
minimizes the quantity~\eqref{eq:(3.3)}.

\section{Numerical experiment for $D=2$ and~$N=1$}
\label{sec:3}
Let us examine Eq.~\eqref{eq:(2.12)} for the field theoretical case,
\begin{equation}
   D=2,\qquad N=1.
\label{eq:(3.1)}
\end{equation}
In Appendix~\ref{sec:A}, we examine Eq.~\eqref{eq:(2.12)} for the quantum
mechanical case, $D=1$ and~$N=1$ following the same method as in the present
section. In the normal ordering scheme, the perturbative coefficients
in~Eq.~\eqref{eq:(2.1)} are given by $c_0=c_1=0$ and~\cite{Serone:2018gjo},
\begin{align}
   c_2&=-\frac{21\zeta(3)}{16\pi^3}\left(\frac{1}{4!}\right)^2
\notag\\
   c_3&=-\frac{27\zeta(3)}{8\pi^4}\left(\frac{1}{4!}\right)^3
\notag\\
   c_4&=-0.116125964(91)\left(\frac{1}{4!}\right)^4
\notag\\
   c_5&=-0.3949534(18)\left(\frac{1}{4!}\right)^5
\notag\\
   c_6&=-1.629794(22)\left(\frac{1}{4!}\right)^6
\notag\\
   c_7&=-7.85404(21)\left(\frac{1}{4!}\right)^7
\notag\\
   c_8&=-43.1920(21)\left(\frac{1}{4!}\right)^8,
\label{eq:(3.2)}
\end{align}
where digits in parentheses are numerical errors associated with the numerical
integration. The first three coefficients, $c_2$, $c_3$, and~$c_4$, coincide
with the coefficients obtained in~Ref.~\cite{Suzuki:1997hb} for~$N=1$ within
the numerical error.

\begin{figure}[htbp]
\centering
\includegraphics[width=0.7\textwidth]{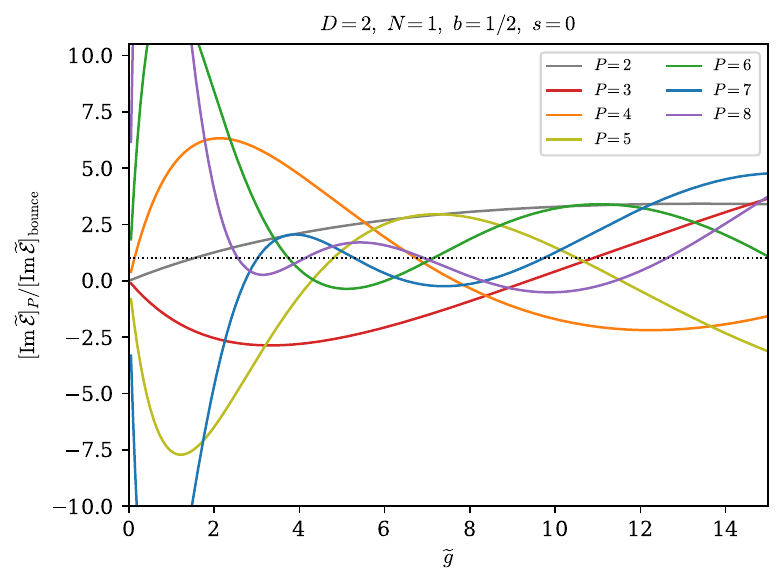}
\caption{The ratio of~$[\imag\widetilde{\mathcal{E}}(\widetilde{g})]_P(b,s)$
obtained by~Eq.~\eqref{eq:(2.12)} and the leading bounce
result~\eqref{eq:(1.2)} as a function of~$\widetilde{g}$ for each
approximation order, $P=2$--$8$; the parameters are fixed
as~$b=(D+N)/2-1=1/2$ and~$s=0$ as in~Ref.~\cite{Suzuki:1997hb}.}
\label{fig:1}
\end{figure}

\begin{figure}[htbp]
\centering
\includegraphics[width=0.7\textwidth]{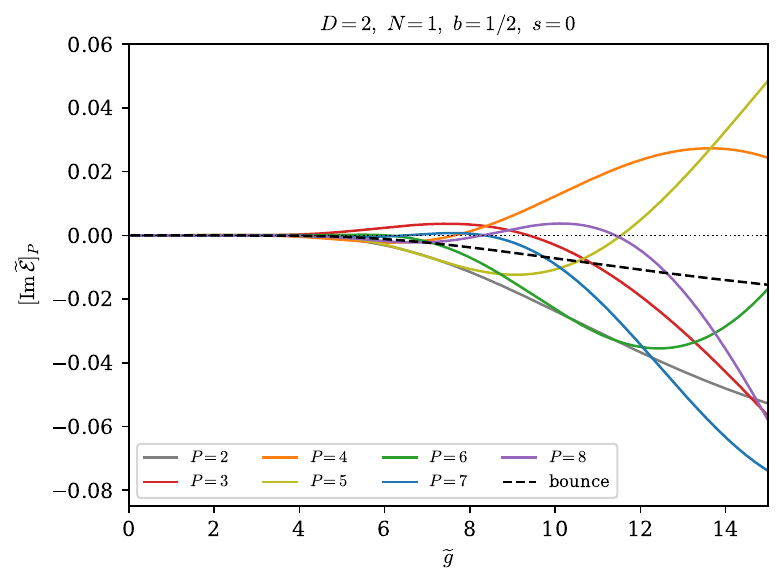}
\caption{Same as~Fig.~\ref{fig:1} but for the value
of~$[\imag\widetilde{\mathcal{E}}(\widetilde{g})]_P(b,s)$ itself; the
parameters are fixed as~$b=(D+N)/2-1=1/2$ and~$s=0$
as in~Ref.~\cite{Suzuki:1997hb}.}
\label{fig:2}
\end{figure}

In~Fig.~\ref{fig:1}, we plot the ratio
of~$[\imag\widetilde{\mathcal{E}}(\widetilde{g})]_P(b,s)$ obtained
by~Eq.~\eqref{eq:(2.12)} and the result of the leading bounce
calculus~\eqref{eq:(1.2)} as a function of~$\widetilde{g}$, simply fixing
$b=(D+N)/2-1=1/2$ and~$s=0$, the same setting as in~Ref.~\cite{Suzuki:1997hb}.
The approximation orders are $P=2$--$8$. The horizontal dotted line is the
value of the leading bounce calculus.
Curves for $P=2$, $3$, and~$4$ reproduce well those of~Fig.~5
of~Ref.~\cite{Suzuki:1997hb}\footnote{%
The overall amplitude is, however, different from that
of~Ref.~\cite{Suzuki:1997hb} because Ref.~\cite{Suzuki:1997hb} refers to the
one-loop determinant of~Ref.~\cite{Brezin:1978} whereas in the present paper we
refer to that of~Ref.~\cite{Malatesta:2017}.} and no sign of convergence is
observed; even the sign of the curves is not definite. Figure~\ref{fig:2} is
the same as~Fig.~\ref{fig:1} but for the value
of~$[\imag\widetilde{\mathcal{E}}(\widetilde{g})]_P(b,s)$ itself.

\begin{figure}[htbp]
\centering
\includegraphics[width=0.7\textwidth]{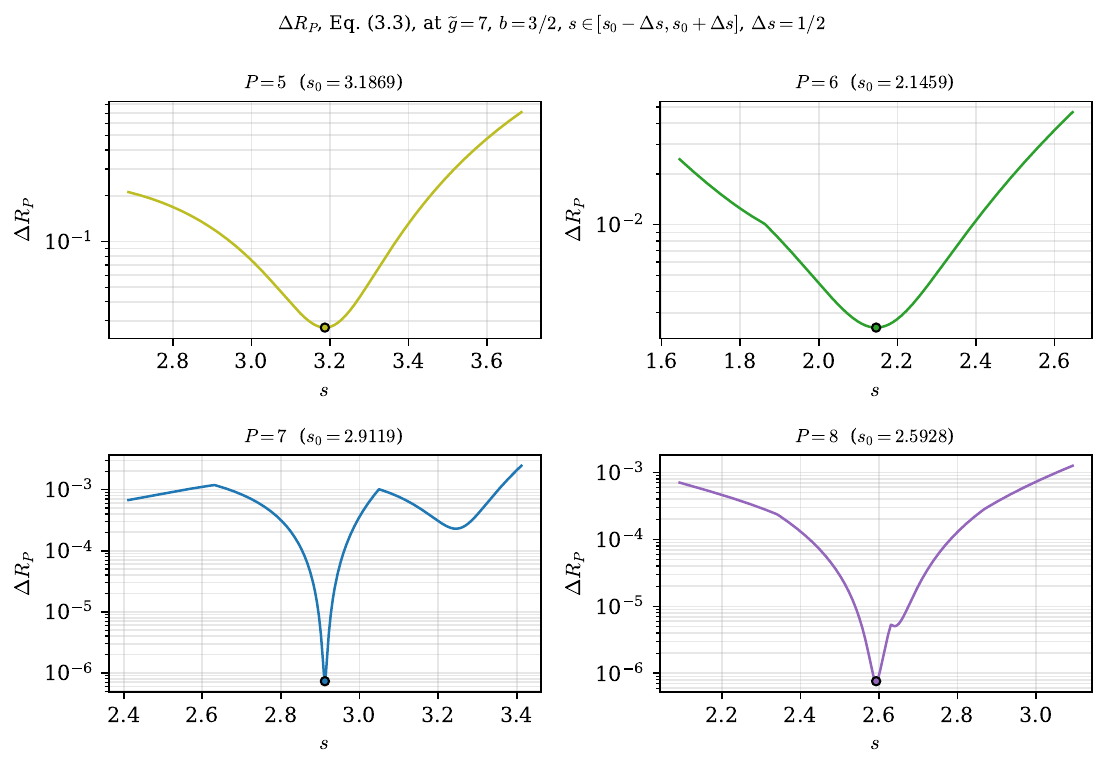}
\caption{$\Delta R_P$~\eqref{eq:(3.3)} as a function of~$s$.}
\label{fig:3}
\end{figure}

Now, we set the value of~$b$ by~Eq.~\eqref{eq:(2.15)}, $b=(D+N)/2=3/2$ and
tune the value of the parameter~$s$ in~Eq.~\eqref{eq:(2.12)}. In this paper,
we choose the parameter~$s$, imitating the strategy
in~Ref.~\cite{Serone:2018gjo}, so that it minimizes the following quantity,
defined in terms of the ratio
$R_P(b,s):=[\imag\widetilde{\mathcal{E}}]_P(b,s)/
[\imag\widetilde{\mathcal{E}}]_{\text{bounce}}$:\footnote{%
The second term of Eq.~\eqref{eq:(3.3)} favors a regular decrease of the
successive differences rather than their smallness, and it alone could vanish
even when both differences are large. In practice, however, the differences
themselves are small at the minimum ($|R_8-R_7|\lesssim2\%$
for~$\widetilde{g}\geq3$), and they are included in the error
estimate~\eqref{eq:(3.4)}. We have also determined $s_0$ by the more direct
criterion $(\partial_sR_P)^2+(R_P-R_{P-1})^2$; it gives $s_0=2.568$ for~$P=8$, and
$R_8$ changes by less than $3\times10^{-4}$ for~$1\leq\widetilde{g}\leq15$.}
\begin{equation}
  \Delta R_P(b,s)
  :=\left\{\partial_sR_P(b,s)\right\}^2
  +\left\{\left|R_P(b,s)-R_{P-1}(b,s)\right|
         -\left|R_{P-1}(b,s)-R_{P-2}(b,s)\right|\right\}^2 .
\label{eq:(3.3)}
\end{equation}
Since $[\imag\widetilde{\mathcal{E}}]_{\text{bounce}}$ does not depend on~$s$, the
minimizing $s$ is the same as that for the corresponding quantity defined
with~$[\imag\widetilde{\mathcal{E}}]_P$. The idea is that the first term of
this quantity accounts for the sensitivity to the choice of~$s$ and the second
term accounts for the dependence on the approximation order.\footnote{%
Since in this paper we fix the value of~$b$ as~Eq.~\eqref{eq:(2.15)} from the
matching with the semiclassical approximation, we have omitted the
term~$\partial_bR_P(b,s)$ from the corresponding criterion
in~Ref.~\cite{Serone:2018gjo}.} In~Fig.~\ref{fig:3}, we plot $\Delta R_P(b,s)$
in~Eq.~\eqref{eq:(3.3)} as a function of~$s$ for each $P=5$, $6$, $7$, and~$8$.
Here, we take $\widetilde{g}=7$ as a representative value of~$\widetilde{g}$;
for other $\widetilde{g}$, the behavior is similar, as~Table~\ref{table:1}
indicates.

\begin{table}[htbp]
\centering
\caption{The parameter $s_0$ that minimizes $\Delta R_P$~\eqref{eq:(3.3)}
with~$b=3/2$. The attained values of~$\Delta R_P$ are also indicated. The
search region is $-2\leq s\leq5$.}
\label{table:1}
\begin{tabular}{c|cc|cc|cc|cc}
\hline\hline
 & \multicolumn{2}{c|}{$P=5$} & \multicolumn{2}{c|}{$P=6$} & \multicolumn{2}{c|}{$P=7$} & \multicolumn{2}{c}{$P=8$}\\
$\widetilde{g}$ & $s_0$ & $\Delta R_P$ & $s_0$ & $\Delta R_P$ & $s_0$ & $\Delta R_P$ & $s_0$ & $\Delta R_P$\\
\hline
1 & 5.000 & $4.2\times 10^{-2}$ & 2.431 & $6.3\times 10^{-2}$ & 3.030 & $4.9\times 10^{-2}$ & 2.523 & $1.0\times 10^{-4}$\\
2 & 2.275 & $6.4\times 10^{-2}$ & 2.226 & $3.5\times 10^{-2}$ & 2.318 & $1.9\times 10^{-2}$ & 2.543 & $1.6\times 10^{-6}$\\
3 & 2.435 & $6.9\times 10^{-2}$ & 2.068 & $1.1\times 10^{-2}$ & 2.081 & $4.4\times 10^{-3}$ & 2.590 & $1.6\times 10^{-7}$\\
4 & 2.675 & $6.9\times 10^{-2}$ & 1.742 & $2.4\times 10^{-7}$ & 2.217 & $3.4\times 10^{-3}$ & 2.525 & $1.1\times 10^{-6}$\\
5 & 2.885 & $6.1\times 10^{-2}$ & 1.941 & $3.7\times 10^{-3}$ & 0.829 & $1.6\times 10^{-4}$ & 2.519 & $1.7\times 10^{-7}$\\
6 & 3.052 & $4.7\times 10^{-2}$ & 2.027 & $4.5\times 10^{-3}$ & 1.829 & $6.0\times 10^{-5}$ & 2.589 & $5.7\times 10^{-7}$\\
7 & 3.187 & $2.7\times 10^{-2}$ & 2.146 & $2.4\times 10^{-3}$ & 2.912 & $7.3\times 10^{-7}$ & 2.593 & $7.6\times 10^{-7}$\\
8 & 3.295 & $7.1\times 10^{-3}$ & 2.326 & $1.2\times 10^{-3}$ & 3.189 & $5.7\times 10^{-9}$ & 2.543 & $2.8\times 10^{-6}$\\
9 & 3.383 & $1.1\times 10^{-3}$ & 2.088 & $1.5\times 10^{-3}$ & 3.405 & $1.0\times 10^{-4}$ & 2.324 & $5.5\times 10^{-6}$\\
10 & 3.453 & $4.1\times 10^{-2}$ & 2.243 & $2.1\times 10^{-4}$ & 3.572 & $1.7\times 10^{-3}$ & 2.438 & $1.3\times 10^{-6}$\\
11 & 1.686 & $4.7\times 10^{-2}$ & 2.460 & $4.8\times 10^{-4}$ & 2.162 & $5.5\times 10^{-3}$ & 3.046 & $2.1\times 10^{-6}$\\
12 & 1.721 & $2.1\times 10^{-2}$ & 2.608 & $1.9\times 10^{-3}$ & 2.144 & $3.0\times 10^{-3}$ & 2.546 & $1.7\times 10^{-7}$\\
13 & 1.774 & $7.9\times 10^{-3}$ & 2.720 & $6.6\times 10^{-3}$ & 2.223 & $1.2\times 10^{-3}$ & 2.574 & $4.2\times 10^{-6}$\\
14 & 1.829 & $2.1\times 10^{-3}$ & 2.811 & $1.9\times 10^{-2}$ & 2.311 & $2.8\times 10^{-4}$ & 2.596 & $3.1\times 10^{-5}$\\
15 & 1.881 & $2.1\times 10^{-4}$ & 2.885 & $4.9\times 10^{-2}$ & 2.389 & $5.5\times 10^{-9}$ & 3.550 & $4.8\times 10^{-6}$\\
\hline\hline
\end{tabular}
\end{table}

In evaluating $[\imag\widetilde{\mathcal{E}}]_P(b,s)$ in what follows, we adopt
the value $s=s_0$ that minimizes~$\Delta R_P(b,s)$~\eqref{eq:(3.3)}
at~$\widetilde{g}=7$ for each~$P$. That is, we employ $s_0=3.1869$ for~$P=5$,
$s_0=2.1459$ for~$P=6$, $s_0=2.9119$ for~$P=7$, $s_0=2.5928$ for~$P=8$, for
all~$\widetilde{g}$. The final results do not depend much on this choice. To
quantify the error associated with this choice of~$s$, we again follow the
strategy of~Ref.~\cite{Serone:2018gjo}. We generate $K$ random~$s$, $s_k$
($k=1$--$K$), uniformly in~$s\in[s_0-\Delta s,s_0+\Delta s]$ and compute
\begin{align}
   \text{Err}\left[\imag\widetilde{\mathcal{E}}\right]_P(b,s_0)
   &:=\frac{1}{\Delta s}
   \frac{1}{K}\sum_{k=1}^K
   \left|\left[\imag\widetilde{\mathcal{E}}\right]_P(b,s_k)
   -\left[\imag\widetilde{\mathcal{E}}\right]_P(b,s_0)\right|
\notag\\
   &\qquad{}
   +\left|\left[\imag\widetilde{\mathcal{E}}\right]_P(b,s_0)
   -\left[\imag\widetilde{\mathcal{E}}\right]_{P-1}(b,s_0)\right|.
\label{eq:(3.4)}
\end{align}
We take $\Delta s=1/2$ and~$K=200$ (this is the same choice as
in~Ref.~\cite{Serone:2018gjo}). Figure~\ref{fig:3} and~Table~\ref{table:1}
indicate that this method with~$\Delta s=1/2$ rather overestimates the error
associated with the choice of~$s$. See also Appendix~\ref{sec:A} on this point.

One might think that, from the asymptotic behavior in~Eq.~\eqref{eq:(2.14)},
the optimal~$s$ can be determined by the strong coupling limit
of~$\imag\widetilde{\mathcal{E}}(\widetilde{g})$. Going back
to~Eq.~\eqref{eq:(1.1)}, when $g$ is very large, $m$ is irrelevant. Since
$[g]=M^{4-D}$ and~$[\mathcal{E}]=M^D$, we would infer that
$\mathcal{E}\sim g^{D/(4-D)}$ for very large $g$ and
$\imag\widetilde{\mathcal{E}}(\widetilde{g})\sim\widetilde{g}^{D/(4-D)}$
as~$\widetilde{g}\gg1$. For~$D=2$, this becomes the strong coupling
limit~$\imag\widetilde{\mathcal{E}}(\widetilde{g})\sim\widetilde{g}$. Then the
optimal $s$ would be~$s_0=1$. However, $s_0$ determined in the above way stays
around~$2$--$3$ as Fig.~\ref{fig:3} shows and this is definitely different
from~$s_0=1$. Note also that the optimal $s_0$ adopted
in~Ref.~\cite{Serone:2018gjo} from a criterion similar to ours, $s_0=5/2$
for~$P=6$, $7$, and~$8$, is quite close to our values
of~$s_0\sim2$--$3$.\footnote{%
On the other hand, values of~$b$ adopted in~Ref.~\cite{Serone:2018gjo},
$b=9$, $13/4$, $11/2$, and~$37/4$, for $P=5$, $6$, $7$, and~$8$, respectively,
are rather larger than our~$b=3/2$. This point could be explained from the
fact that their criterion that corresponds to~Eq.~\eqref{eq:(3.3)} is rather
insensitive to~$b$.} One possible reason for the above discrepancy from the
expectation~$s=1$ is that, in~$D=2$, the mass counterterm behaves
as~$g^2\ln(\Lambda/m)$, where $\Lambda$ is the ultraviolet cutoff, and the
massless limit is ill defined. Then the above simple dimensional argument
breaks down and logarithmic corrections such
as~$\imag\widetilde{\mathcal{E}}\sim\widetilde{g}(\ln\widetilde{g})^\alpha$ with
some exponent~$\alpha$ may arise. If one approximates this behavior by a
simple power~$\widetilde{g}^s$, $s$ would appear rather different
from~$1$.\footnote{%
For $D=1$, the dimensional argument
predicts~$\imag\widetilde{\mathcal{E}}\sim\widetilde{g}^{1/3}$, i.e.\ $s=1/3$.
As we see in~Appendix~\ref{sec:A}, the optimal $s_0$ in~$D=1$ is $0.2$--$0.4$
for~$P\geq7$ and this is very close to~$1/3$. This fact supports this reasoning
for the $D=2$ case.} Another possible reason is that with the presently
available perturbative coefficients the parameter~$s$ improves the behavior in
the intermediate region of~$\widetilde{g}$ rather than realizing the true
strong coupling limit. For these reasons, we adopt the values of~$s_0$ quoted
above.

\begin{figure}[htbp]
\centering
\includegraphics[width=0.7\textwidth]{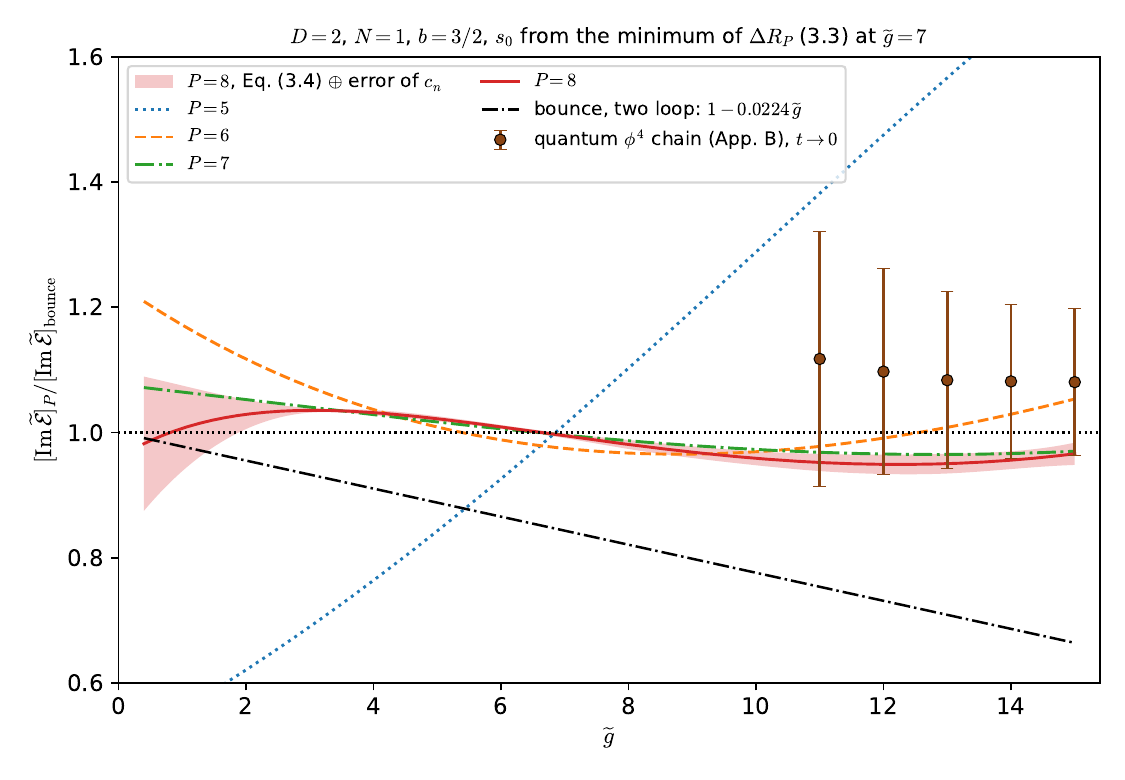}
\caption{The ratio of~$[\imag\widetilde{\mathcal{E}}(\widetilde{g})]_P(b,s_0)$
obtained by~Eq.~\eqref{eq:(2.12)} and the leading bounce
result~\eqref{eq:(1.2)} as a function of~$\widetilde{g}$ for approximation
orders~$P=5$--$8$; the parameters are taken to be $b=3/2$ and~$s=s_0$ which
minimizes the quantity in~Eq.~\eqref{eq:(3.3)}. The error band for~$P=8$ is
estimated as explained in the text. We have also shown the result of the
quantum $\phi^4$ chain in~Appendix~\ref{sec:B}.}
\label{fig:4}
\end{figure}

Now, in~Fig.~\ref{fig:4}, we plot the ratio
of~$[\imag\widetilde{\mathcal{E}}(\widetilde{g})]_P(b,s_0)$ obtained
by~Eq.~\eqref{eq:(2.12)} and the result of the leading bounce
calculus~\eqref{eq:(1.2)} as a function of~$\widetilde{g}$. The horizontal
dotted line is the value of the leading bounce calculus. The parameters are
taken to be~$b=3/2$ and $s=s_0$ which minimizes the quantity
in~Eq.~\eqref{eq:(3.3)}. The error for~$P=8$ is shown as a band. For
completeness, we have taken into account not only the error estimated
by~Eq.~\eqref{eq:(3.4)} but also the error associated with the
coefficients~$c_n$ in~Eq.~\eqref{eq:(3.2)} in quadrature; the uncertainty
induced by the latter is, however, an order of magnitude smaller than that from
the former. We also plot the bounce result with the two-loop correction (the
dash-dotted line); according to~Ref.~\cite{Malatesta:2017}, the leading bounce
result~\eqref{eq:(1.2)} is multiplied by the factor,
\begin{equation}
   1+\left[\xi-\frac{4}{3}VG_0(0)^2
   -3I_2K(0)\right]\left(\frac{-\widetilde{g}}{3!}\right)
   =1-0.0224\widetilde{g},
\label{eq:(3.5)}
\end{equation}
where we have used the abbreviations in~Ref.~\cite{Malatesta:2017};
$I_2=(4-D)S_0/3!$ and, for~$D=2$, $\xi-\frac{4}{3}VG_0(0)^2=0.6553(2)$
and~$K(0)=0.0148430$. The factor~$-3I_2K(0)=-S_0K(0)$ arises from the
translation of the (finite) mass renormalization scheme from that
in~Ref.~\cite{Malatesta:2017} to the present normal ordering scheme.
See~Eq.~(69) of~Ref.~\cite{Malatesta:2017}.

Remarkably, in~Fig.~\ref{fig:4}, the curves of~$P=6$, $P=7$ and~$P=8$ show a
clear sign of convergence; for~$\widetilde{g}\gtrsim4$, the $P=7$ and~$P=8$
results agree within~$1$--$2\%$, comparable to the estimated error of~$P=8$.
Compare this plot with~Fig.~\ref{fig:1}, which shows a violently oscillating
behavior. We thus see that an appropriate choice of the free parameters $b$
and~$s$ can drastically change the situation; it is actually possible to
extract the imaginary part of the vacuum energy density in field theory (at
least for the present $D=2$, $N=1$ case) from conventional perturbation theory.
In~Table~\ref{table:2}, we tabulate the values of the ratio with estimated
errors.
\begin{table}[htbp]
\centering
\caption{The
ratio~$[\imag\widetilde{\mathcal{E}}]_P(b,s_0)/[\imag\widetilde{\mathcal{E}}]_{\text{bounce}}$ with~$b=3/2$. For each~$P$, $s_0$ is fixed to the value that
minimizes Eq.~\eqref{eq:(3.3)} at~$\widetilde{g}=7$. The number in the
parentheses is the total error, i.e.\ the error of Eq.~\eqref{eq:(3.4)}
with~$\Delta s=1/2$ and~$K=200$ combined in quadrature with the uncertainty
induced by the errors of the coefficients~\eqref{eq:(3.2)}, the latter being
estimated with $400$ Gaussian samples of~$c_n$.}
\label{table:2}
\begin{tabular}{c|cccc}
\hline\hline
$\widetilde{g}$ & $P=5$ & $P=6$ & $P=7$ & $P=8$\\
\hline
1 & $0.5573 \pm 0.6551$ & $1.1717 \pm 0.4319$ & $1.0646(949)$ & $1.0066(684)$\\
2 & $0.6208 \pm 0.6867$ & $1.1174 \pm 0.2709$ & $1.0525(588)$ & $1.0289(235)$\\
3 & $0.6897 \pm 0.7113$ & $1.0726 \pm 0.1499$ & $1.0403(301)$ & $1.0354(37)$\\
4 & $0.7637 \pm 0.7263$ & $1.0367(612)$ & $1.0283(101)$ & $1.0317(40)$\\
5 & $0.8426 \pm 0.7291$ & $1.0087(319)$ & $1.0167(72)$ & $1.0220(40)$\\
6 & $0.9257 \pm 0.7169$ & $0.9882(411)$ & $1.0058(167)$ & $1.0089(30)$\\
7 & $1.0126 \pm 0.6867$ & $0.9745(632)$ & $0.9957(207)$ & $0.9947(37)$\\
8 & $1.1025 \pm 0.6599$ & $0.9671(703)$ & $0.9868(215)$ & $0.9809(72)$\\
9 & $1.1947 \pm 0.6736$ & $0.9655(688)$ & $0.9791(219)$ & $0.9686(93)$\\
10 & $1.2882 \pm 0.7409$ & $0.9692(784)$ & $0.9729(191)$ & $0.9588(111)$\\
11 & $1.3820 \pm 0.8080$ & $0.9778(934)$ & $0.9683(128)$ & $0.9521(137)$\\
12 & $1.4752 \pm 0.8702$ & $0.9909 \pm 0.1036$ & $0.9656(246)$ & $0.9491(154)$\\
13 & $1.5665 \pm 0.9249$ & $1.0081 \pm 0.1083$ & $0.9649(430)$ & $0.9502(157)$\\
14 & $1.6546 \pm 0.9694$ & $1.0291 \pm 0.1069$ & $0.9663(633)$ & $0.9558(142)$\\
15 & $1.7382 \pm 1.0005$ & $1.0534 \pm 0.1203$ & $0.9701(849)$ & $0.9660(178)$\\
\hline\hline
\end{tabular}
\end{table}
The results are insensitive to the reference coupling used to fix $s_0$. If
$s_0$ is instead determined at~$\widetilde{g}=4$ or~$\widetilde{g}=10$
($s_0=2.525$ and $2.438$ for~$P=8$), or at each $\widetilde{g}$ separately
(Table~1), $R_8$ changes by less than~$0.002$ for~$\widetilde{g}\leq14$;
at~$\widetilde{g}=15$, the minimum of~Eq.~\eqref{eq:(3.3)} jumps to another
local minimum, $s_0=3.550$, which gives $R_8=0.931$. On the other hand, with
$s=1$, the value suggested by the strong coupling limit, the results do not
converge in~$P$; e.g., $R_P=1.48$, $0.87$, $0.72$ and~$1.04$
for $P=5$--$8$ at~$\widetilde{g}=7$.

The present approach finally yields, in the intermediate coupling region, an
imaginary part rather close to the leading-order semiclassical approximation
with the one-loop determinant~\eqref{eq:(1.2)}. On the other hand, somewhat
unexpectedly, the imaginary part is larger by~$10$--$20\%$ than the
semiclassical result with the two-loop correction computed
in~Ref.~\cite{Malatesta:2017}. If the present resummation is taken at face
value, this would imply that the higher-order corrections in the semiclassical
approximation are not small in the intermediate region and almost compensate
the two-loop correction. We note, however, that the error
estimate~\eqref{eq:(3.4)} does not include a possible systematic uncertainty of
the resummation itself; see Section~\ref{sec:4} and Appendix~\ref{sec:B}.

\begin{figure}[htbp]
\centering
\includegraphics[width=0.7\textwidth]{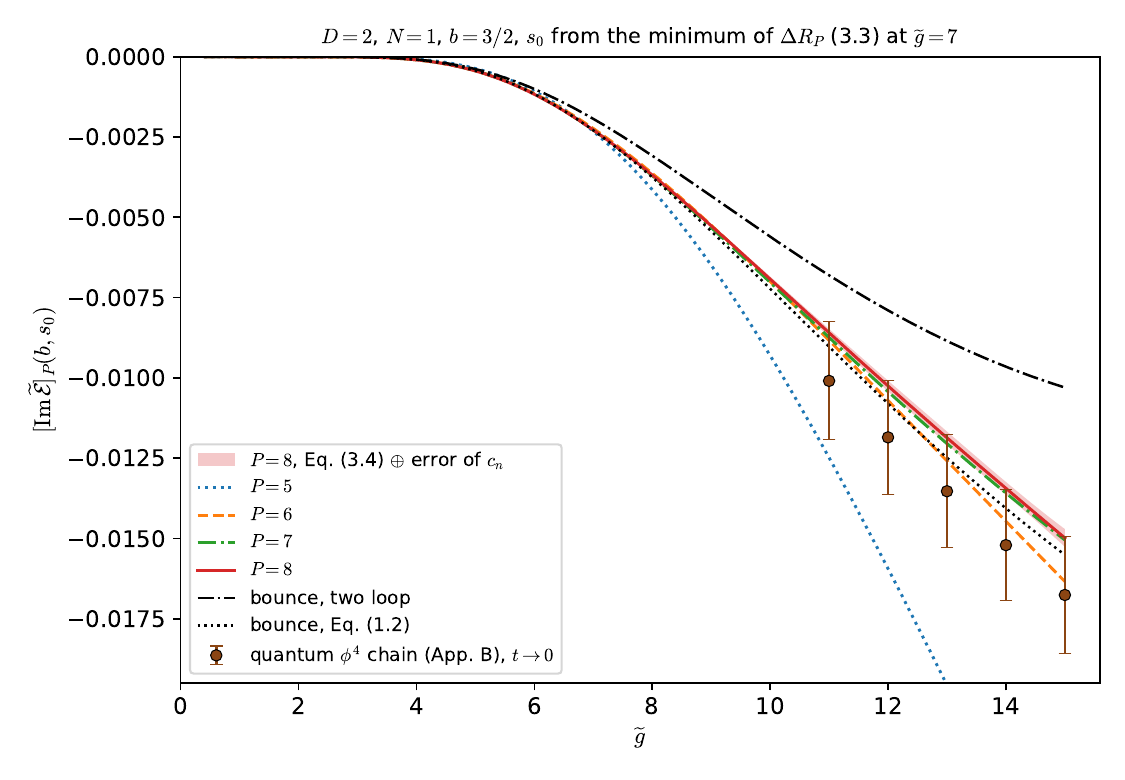}
\caption{Same as~Fig.~\ref{fig:4} but for the value of
$[\imag\widetilde{\mathcal{E}}(\widetilde{g})]_P(b,s)$ itself; the parameters
are fixed as~$b=(D+N)/2=3/2$ and~$s=s_0$, which minimizes the quantity
in~Eq.~\eqref{eq:(3.3)}. The error band for~$P=8$ is estimated as explained in
the text. We have also shown the result of the quantum $\phi^4$ chain
in~Appendix~\ref{sec:B}.}
\label{fig:5}
\end{figure}

Figure~\ref{fig:5} is the same as~Fig.~\ref{fig:4} but for the value
of~$[\imag\widetilde{\mathcal{E}}(\widetilde{g})]_P(b,s)$ itself. We also
tabulate the values of~$[\imag\widetilde{\mathcal{E}}(\widetilde{g})]_P(b,s)$
in~Table~\ref{table:3}.

\begin{table}[htbp]
\centering
\caption{The same as Table~\ref{table:2} but for the value
of~$[\imag\widetilde{\mathcal{E}}]_P(b,s_0)$ itself. The last column is the
semiclassical result~\eqref{eq:(1.2)}.}
\label{table:3}
\scriptsize
\begin{tabular}{c|cccc|c}
\hline\hline
$\widetilde{g}$ & $P=5$ & $P=6$ & $P=7$ & $P=8$ & bounce\\
\hline
1 & $(-7.660 \pm 9.006)\times 10^{-16}$ & $-1.611(594)\times 10^{-15}$ & $-1.463(130)\times 10^{-15}$ & $-1.384(94)\times 10^{-15}$ & $-1.375\times 10^{-15}$\\
2 & $(-1.789 \pm 1.979)\times 10^{-8}$ & $-3.220(780)\times 10^{-8}$ & $-3.033(170)\times 10^{-8}$ & $-2.965(68)\times 10^{-8}$ & $-2.881\times 10^{-8}$\\
3 & $(-4.602 \pm 4.747)\times 10^{-6}$ & $(-7.158 \pm 1.000)\times 10^{-6}$ & $-6.942(201)\times 10^{-6}$ & $-6.909(25)\times 10^{-6}$ & $-6.673\times 10^{-6}$\\
4 & $(-7.124 \pm 6.775)\times 10^{-5}$ & $-9.670(571)\times 10^{-5}$ & $-9.593(94)\times 10^{-5}$ & $-9.624(37)\times 10^{-5}$ & $-9.328\times 10^{-5}$\\
5 & $(-3.637 \pm 3.147)\times 10^{-4}$ & $-4.354(138)\times 10^{-4}$ & $-4.389(31)\times 10^{-4}$ & $-4.411(17)\times 10^{-4}$ & $-4.317\times 10^{-4}$\\
6 & $-1.073(831)\times 10^{-3}$ & $-1.145(48)\times 10^{-3}$ & $-1.166(19)\times 10^{-3}$ & $-1.169(4)\times 10^{-3}$ & $-1.159\times 10^{-3}$\\
7 & $(-2.321 \pm 1.574)\times 10^{-3}$ & $-2.233(145)\times 10^{-3}$ & $-2.282(47)\times 10^{-3}$ & $-2.280(8)\times 10^{-3}$ & $-2.292\times 10^{-3}$\\
8 & $(-4.138 \pm 2.477)\times 10^{-3}$ & $-3.630(264)\times 10^{-3}$ & $-3.703(81)\times 10^{-3}$ & $-3.681(27)\times 10^{-3}$ & $-3.753\times 10^{-3}$\\
9 & $(-6.490 \pm 3.659)\times 10^{-3}$ & $-5.245(374)\times 10^{-3}$ & $-5.319(119)\times 10^{-3}$ & $-5.262(51)\times 10^{-3}$ & $-5.432\times 10^{-3}$\\
10 & $(-9.302 \pm 5.350)\times 10^{-3}$ & $-6.999(566)\times 10^{-3}$ & $-7.025(138)\times 10^{-3}$ & $-6.923(80)\times 10^{-3}$ & $-7.221\times 10^{-3}$\\
11 & $-1.248(730)\times 10^{-2}$ & $-8.832(844)\times 10^{-3}$ & $-8.746(116)\times 10^{-3}$ & $-8.600(124)\times 10^{-3}$ & $-9.032\times 10^{-3}$\\
12 & $-1.594(940)\times 10^{-2}$ & $-1.070(112)\times 10^{-2}$ & $-1.043(27)\times 10^{-2}$ & $-1.025(17)\times 10^{-2}$ & $-1.080\times 10^{-2}$\\
13 & $(-1.956 \pm 1.155)\times 10^{-2}$ & $-1.259(135)\times 10^{-2}$ & $-1.205(54)\times 10^{-2}$ & $-1.187(20)\times 10^{-2}$ & $-1.249\times 10^{-2}$\\
14 & $(-2.327 \pm 1.363)\times 10^{-2}$ & $-1.447(150)\times 10^{-2}$ & $-1.359(89)\times 10^{-2}$ & $-1.344(20)\times 10^{-2}$ & $-1.406\times 10^{-2}$\\
15 & $(-2.696 \pm 1.552)\times 10^{-2}$ & $-1.634(187)\times 10^{-2}$ & $-1.505(132)\times 10^{-2}$ & $-1.498(28)\times 10^{-2}$ & $-1.551\times 10^{-2}$\\
\hline\hline
\end{tabular}
\end{table}

\clearpage

\section{Conclusion}
\label{sec:4}
We have revisited the approach of~Refs.~\cite{Suzuki:1996rt,Suzuki:1997hb},
which extracts the imaginary part of the vacuum energy density from
conventional perturbative coefficients. With the first seven nontrivial
coefficients of~Ref.~\cite{Serone:2018gjo} and the two tunable parameters $b$
and~$s$ in the finite-order truncated Borel transform, the violently
oscillating behavior of~Fig.~\ref{fig:1} turns into the convergent one
of~Fig.~\ref{fig:4}: for $\widetilde g\gtrsim4$, the $P=7$ and~$P=8$ results
agree to within~$1$--$2\%$. It is thus possible, at least for $D=2$ and~$N=1$,
to extract this nonperturbative quantity in field theory from conventional
perturbation theory. The quantum mechanical case in~Appendix~\ref{sec:A}, where
the exact result is known, provides a consistency check of the same strategy.

The resulting imaginary part is close to the leading-order semiclassical
approximation~\eqref{eq:(1.2)}\footnote{%
The normalization of~Eq.~\eqref{eq:(1.2)} is controlled by~$C_{2,1}$, for which
the values of Refs.~\cite{Brezin:1978} and~\cite{Malatesta:2017} differ by more
than~$20\%$. Here, we adopt the latter value.} in the intermediate coupling
region, $3\lesssim\widetilde{g}\lesssim8$. On the other hand, it
is~$10$--$20\%$ larger than the semiclassical result including the two-loop
correction~\eqref{eq:(3.5)}. Our error estimate~\eqref{eq:(3.4)} quantifies
only the stability with respect to~$s$ and to the approximation order, not a
systematic uncertainty of the resummation ansatz itself; for $D=1$
in~Appendix~\ref{sec:A}, however, the error~\eqref{eq:(3.4)} always exceeds the
actual deviation from the exact result for~$P\geq7$. Still, if the present
result is taken at face value, it implies that the higher-order corrections
around the bounce are rather large in this region and almost compensate the
two-loop correction. As Figs.~\ref{fig:4} and~\ref{fig:5} show, our result is,
for~$11\leq\widetilde{g}\leq15$, consistent within errors with an independent
nonperturbative computation, the continuum extrapolation of the
Pad\'e-resummed linked-cluster expansion in a quantum $\phi^4$ chain presented
in~Appendix~\ref{sec:B}.

A natural next step is to apply the present strategy to~$D=3$, for which the
perturbative coefficients for the vacuum energy density are available to nine
loops for any~$N$~\cite{Sberveglieri:2020eko}.

\section*{Acknowledgements}
I am grateful to the organizers of the RIMS–iTHEMS joint workshop ``Resurgence
Theory and New nonperturbative Analysis in Quantum Theory'' (Kobe, September
6--8, 2017), especially Tatsuhiro Misumi; encouragement from Jean Zinn-Justin
on that occasion prompted me to return to the present problem.
This work was supported by JSPS KAKENHI Grant Number JP23739313 and by the MEXT
program ``Supporting Pioneering Research through AI for 1,000 Discovery
Challenges (SPReAD),'' Grant Number JPMXP1726257479.
It was also partially supported by the Quantum and Spacetime Research
Institute, Kyushu University.
I received substantial assistance from Claude (Anthropic) in carrying out this
study. Needless to say, any remaining errors are the author's own.

\appendix

\section{Quantum mechanics}
\label{sec:A}
\begin{figure}[htbp]
\centering
\includegraphics[width=0.7\textwidth]{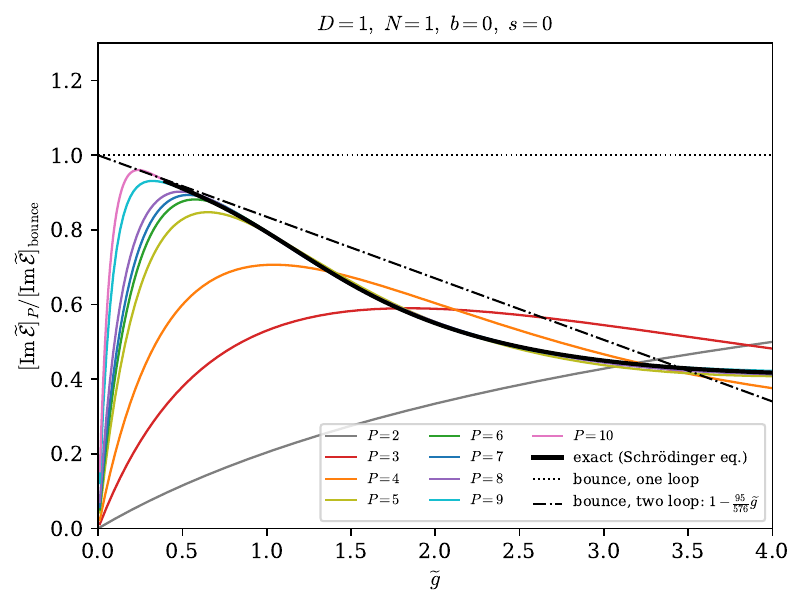}
\caption{The ratio
of~$[\imag\widetilde{\mathcal{E}}(\widetilde{g})]_P(b,s)$~\eqref{eq:(2.12)} and
the leading bounce result~\eqref{eq:(1.2)} as a function of~$\widetilde{g}$
for approximation orders, $P=2$--$10$; the parameters are fixed as~$b=0$
and~$s=0$ as in~Refs.~\cite{Suzuki:1996rt,Suzuki:1997hb}.}
\label{fig:A1}
\end{figure}

In this appendix, to provide further support for the approach in the main text,
we study a quantum mechanical case, $D=1$ and~$N=1$. This system was already
studied in~Refs.~\cite{Suzuki:1996rt,Suzuki:1997hb} with the choice of
parameters, $b=(D+N)/2-1=0$ and~$s=0$ in~Eq.~\eqref{eq:(2.12)}. Here we
examine the effect of the choice, $b=(D+N)/2=1$ and~$s=s_0$, which minimizes
the stability criterion~\eqref{eq:(3.3)}. The error estimation
by~Eq.~\eqref{eq:(3.4)} with $\Delta s=1/2$ and~$K=200$ is also identical to
the main text. The perturbative coefficients in~Eq.~\eqref{eq:(2.1)} can be
obtained by the method in~Ref.~\cite{Bender:1969si}:
\begin{align}
c_0&=\frac{1}{2},\quad
c_1=-\frac{1}{32},\quad
c_2=-\frac{7}{1536},\quad
c_3=-\frac{37}{24576},\quad
c_4=-\frac{10295}{14155776},
\nonumber\\
c_5&=-\frac{33953}{75497472},\quad
c_6=-\frac{21839467}{65229815808},\quad
c_7=-\frac{302588297}{1043677052928},\quad
\nonumber\\
c_8&=-\frac{343498366351}{1202315964973056},\quad
c_9=-\frac{6069664723495}{19237055439568896},
\notag\\
c_{10}&=-\frac{2139002477196865}{5540271966595842048}.
\end{align}
First, in~Fig.~\ref{fig:A1}, we show the result of the earlier choice of
parameters, $b=0$ and~$s=0$ for the ratio to the leading bounce
result~\eqref{eq:(1.2)}. The dash-dotted line shows the result of the bounce
calculus with the two-loop correction~\cite{Zinn-Justin:1979jnt}. The thick
line is the exact value of the imaginary part that is obtained by directly
solving the Schr\"odinger equation with the complex scaling method. As observed
in~Refs.~\cite{Suzuki:1996rt,Suzuki:1997hb}, Eq.~\eqref{eq:(2.12)} shows a
smooth converging behavior; as noted there, the convergence is faster in the
strong coupling region and slower in the weak coupling region. In any case, the
formula works quite well in this quantum mechanics case. In this~$D=1$ case,
the bounce calculus with the two-loop corrections as well as our resummed
result both show that the tunneling rate is suppressed relative to the leading
bounce result.

\begin{figure}[htbp]
\centering
\includegraphics[width=0.7\textwidth]{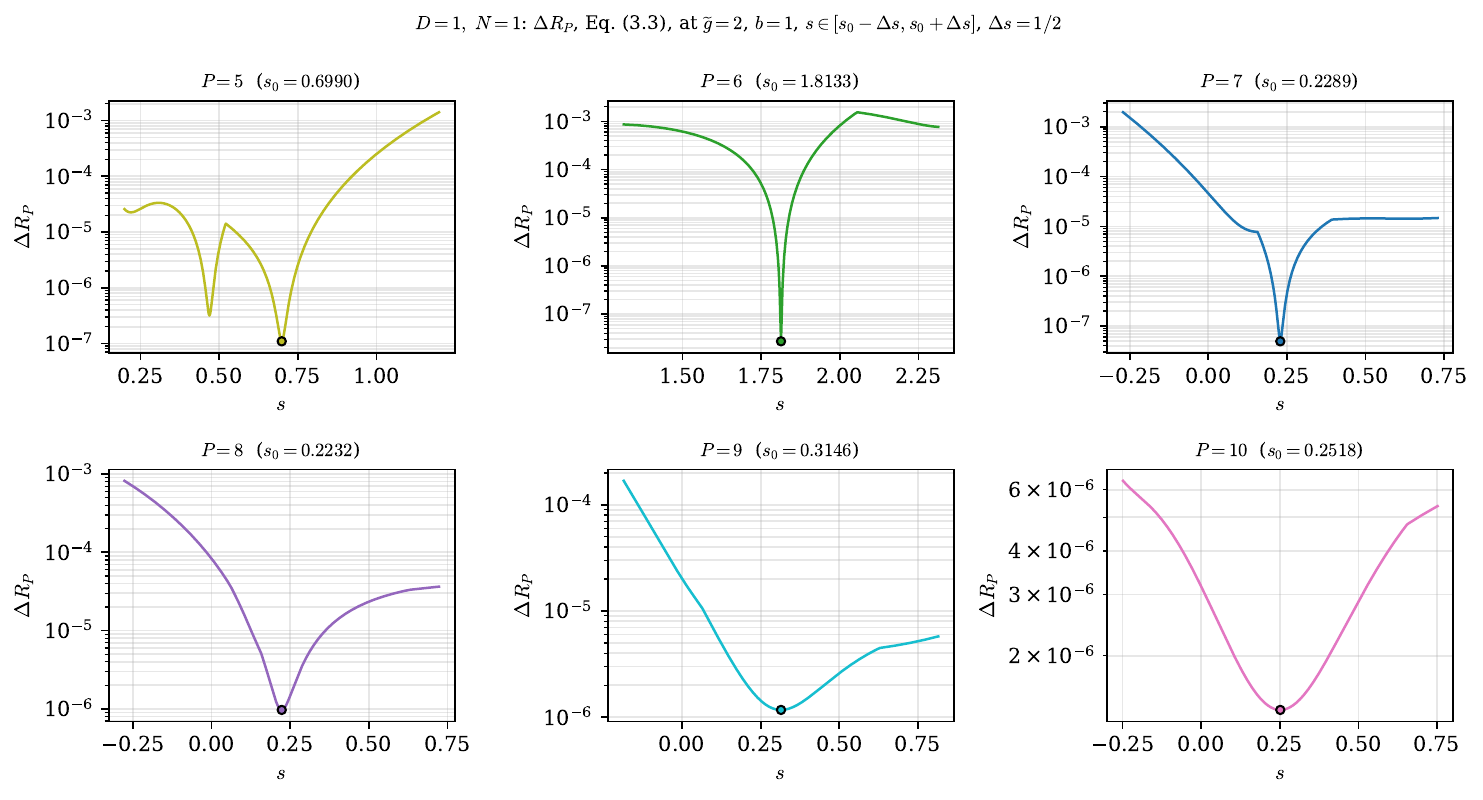}
\caption{$\Delta R_P(b,s)$~\eqref{eq:(3.3)} as a function of~$s$.
$\widetilde{g}=2$.}
\label{fig:A2}
\end{figure}

\begin{figure}[htbp]
\centering
\includegraphics[width=0.7\textwidth]{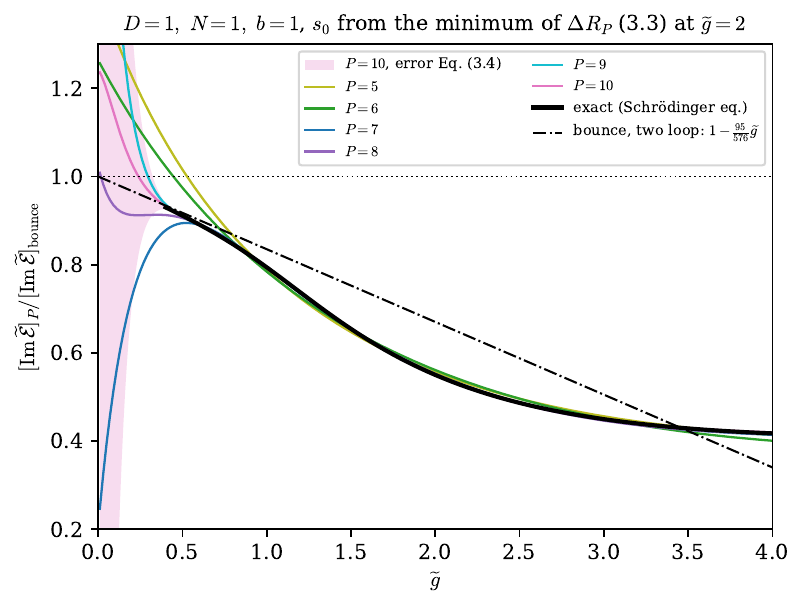}
\caption{The ratio
of~$[\imag\widetilde{\mathcal{E}}(\widetilde{g})]_P(b,s_0)$~\eqref{eq:(2.12)}
and the leading bounce result~\eqref{eq:(1.2)} as a function of~$\widetilde{g}$
for approximation orders~$P=5$--$10$; the parameters are taken to be $b=1$
and~$s=s_0$, which minimizes the quantity in~Eq.~\eqref{eq:(3.3)}. The error
band for~$P=10$ is estimated as explained in the text.}
\label{fig:A3}
\end{figure}

Next, we examine the approach in the present paper, i.e.\ the parameter~$b$ is
fixed by~Eq.~\eqref{eq:(2.15)} and the parameter~$s$ is tuned to the value
that minimizes~Eq.~\eqref{eq:(3.3)}. In Fig.~\ref{fig:A2}, for each~$P$, the
quantity~\eqref{eq:(3.3)} is plotted as a function of~$s$
for~$\widetilde{g}=2$. Quite interestingly, for~$P\geq7$, the optimal value
of~$s$, $s_0$, is $0.2$--$0.4$ and this is close to~$1/3$, the value expected
from the strong coupling behavior
$\imag\widetilde{\mathcal{E}}\sim\widetilde{g}^{1/3}$. Figure~\ref{fig:A3} is
the ratio
of~$[\imag\widetilde{\mathcal{E}}(\widetilde{g})]_P(b,s_0)$~\eqref{eq:(2.12)}
and the leading bounce result~\eqref{eq:(1.2)} as a function
of~$\widetilde{g}$; as $s_0$, we employed values quoted in~Fig.~\ref{fig:A2}
for all~$\widetilde{g}$. We observe that the same strategy adopted in the main
text for~$D=2$ and~$N=1$ also works well for the quantum mechanical case. Note
that although the value of~$b$ in~Eq.~\eqref{eq:(2.15)} is chosen so that the
weak coupling behavior~\eqref{eq:(2.14)} agrees with the semiclassical
result~\eqref{eq:(1.2)}, only the power of~$\widetilde{g}$ is matched
by~Eq.~\eqref{eq:(2.15)}, not the prefactor. Thus, it is not unexpected that
the convergence for the weak coupling~$\widetilde{g}\sim0$ is slow
in~Fig.~\ref{fig:A3}. The ratio is also tabulated in~Table~\ref{table:A1}.

For $P\geq7$, the deviation from the exact result is comparable to or smaller
than the error~\eqref{eq:(3.4)}; the ratio of the former to the latter is at
most about~$1.2$ and typically below~$0.8$, the precise value depending on the
random sample~$s_k$ in~Eq.~\eqref{eq:(3.4)}. For $P\leq6$, the
estimate~\eqref{eq:(3.4)} is not conservative.

\begin{table}[htbp]
\centering
\caption{The ratio $[\imag\widetilde{\mathcal{E}}]_P(b,s_0)/[\imag\widetilde{\mathcal{E}}]_{\text{bounce}}$ with $b=1$. $s_0$~is fixed to the value that minimizes
Eq.~\eqref{eq:(3.3)} at~$\widetilde{g}=2$; see~Fig.~\ref{fig:A2}. The number in
the parentheses is the error of~Eq.~\eqref{eq:(3.4)} with $\Delta s=1/2$
and~$K=200$. The last column is the exact ratio, obtained by solving the
Schr\"odinger equation with the complex scaling method.}
\label{table:A1}
\scriptsize
\begin{tabular}{c|cccccc|c}
\hline\hline
$\widetilde g$ & $P=5$ & $P=6$ & $P=7$ & $P=8$ & $P=9$ & $P=10$ & exact\\
\hline
0.50 & $1.0157(869)$ & $0.9744 \pm 0.1094$ & $0.8942(453)$ & $0.9057(156)$ & $0.9092(45)$ & $0.9099(23)$ & 0.9111\\
0.75 & $0.8872(452)$ & $0.8700(545)$ & $0.8629(56)$ & $0.8605(41)$ & $0.8590(17)$ & $0.8588(8)$ & 0.8582\\
1.00 & $0.7878(196)$ & $0.7841(194)$ & $0.7934(38)$ & $0.7928(18)$ & $0.7939(17)$ & $0.7937(6)$ & 0.7941\\
1.25 & $0.7095(62)$ & $0.7127(41)$ & $0.7200(34)$ & $0.7211(21)$ & $0.7219(13)$ & $0.7222(7)$ & 0.7226\\
1.50 & $0.6471(49)$ & $0.6531(168)$ & $0.6539(39)$ & $0.6551(18)$ & $0.6546(13)$ & $0.6549(7)$ & 0.6543\\
1.75 & $0.5968(95)$ & $0.6030(224)$ & $0.5981(29)$ & $0.5985(20)$ & $0.5972(16)$ & $0.5971(7)$ & 0.5963\\
2.00 & $0.5561(112)$ & $0.5610(230)$ & $0.5525(26)$ & $0.5519(23)$ & $0.5506(20)$ & $0.5501(10)$ & 0.5500\\
2.25 & $0.5230(110)$ & $0.5257(221)$ & $0.5160(44)$ & $0.5146(24)$ & $0.5139(19)$ & $0.5133(10)$ & 0.5141\\
2.50 & $0.4959(103)$ & $0.4960(206)$ & $0.4873(52)$ & $0.4854(23)$ & $0.4857(19)$ & $0.4852(16)$ & 0.4866\\
2.75 & $0.4739(93)$ & $0.4713(169)$ & $0.4649(54)$ & $0.4630(34)$ & $0.4644(26)$ & $0.4642(18)$ & 0.4656\\
3.00 & $0.4561(78)$ & $0.4507(116)$ & $0.4478(56)$ & $0.4463(41)$ & $0.4485(28)$ & $0.4486(17)$ & 0.4497\\
3.25 & $0.4416(58)$ & $0.4339(143)$ & $0.4349(50)$ & $0.4340(42)$ & $0.4368(34)$ & $0.4373(20)$ & 0.4376\\
3.50 & $0.4301(36)$ & $0.4201(246)$ & $0.4254(37)$ & $0.4253(37)$ & $0.4283(44)$ & $0.4293(20)$ & 0.4287\\
3.75 & $0.4209(44)$ & $0.4092(349)$ & $0.4187(37)$ & $0.4195(44)$ & $0.4223(52)$ & $0.4236(20)$ & 0.4221\\
4.00 & $0.4139(73)$ & $0.4007(448)$ & $0.4143(58)$ & $0.4159(48)$ & $0.4182(55)$ & $0.4197(28)$ & 0.4174\\
\hline\hline
\end{tabular}
\end{table}

\clearpage

\section{Quantum $\phi^4$ chain}
\label{sec:B}
In this appendix, we compute the imaginary part of the vacuum energy density
for~$D=2$ and~$N=1$ in an independent way, on the basis of a quantum $\phi^4$
chain on a spatial lattice. Following Ref.~\cite{Nishiyama:2001}, we compute a
strong coupling expansion of the ground-state energy density by the
linked-cluster expansion~\cite{Marland:1981,Hamer:1984,Singh:1988,Oitmaa:2006},
extrapolate it to the weak coupling region by the Pad\'e approximation, and
obtain the imaginary part by the analytic continuation to the unbounded
potential. To control the continuum limit, we also carry out the bounce
calculus on the same lattice. As explained below, the lattice artifacts turn
out to be large, and the present approach does not reach the precision of the
main text; we therefore present only a qualitative conclusion.

\subsection{Lattice Hamiltonian}
We consider the lattice Hamiltonian,
\begin{equation}
   H=\sum_{i\in\mathbb{Z}}
   \left[
   \frac{1}{2}\pi_i^2
   +\frac{a_{\text{I}}}{2}(\phi_i-\phi_{i+1})^2
   +\frac{b_{\text{I}}}{2}(\phi_i-\phi_{i+2})^2
   +\frac{1}{2}r\phi_i^2
   +\rho\phi_i^4
   \right],\qquad
   [\phi_i,\pi_j]=i\delta_{ij},
\label{eq:(B1)}
\end{equation}
where $a_{\text{I}}=1-4b_{\text{I}}$. The choice $b_{\text{I}}=-1/12$ is the
tree-level Symanzik improvement, i.e.\ the dispersion
$\omega_k^2:=4a_{\text{I}}\sin^2(k/2)+4b_{\text{I}}\sin^2k=k^2-k^6/90+\cdots$ has no
$O(k^4)$~term. With $t:=m^2a^2$, where $m$ is identified with the mass parameter
in Eq.~\eqref{eq:(1.1)}, the continuum limit is $t\to0$. The quartic coupling
is~$\rho=-\widetilde{g}t/24$ in the present super-renormalizable theory. For
the mass parameter, the normal ordering scheme in the main text requires
\begin{equation}
   r=t-12\rho G(t),\qquad
   G(t)=\int_{-\pi}^\pi\frac{dk}{2\pi}\,\frac{1}{2\sqrt{\omega_k^2+t}}
   \stackrel{t\to0}{\to}-\frac{1}{4\pi}\ln t+\text{const.}
\label{eq:(B2)}
\end{equation}
The vacuum energy density in the normal ordering scheme is given by
$\widetilde{\mathcal{E}}=[e_{\text{site}}-e_{\text{free}}(t)+3\rho G(t)^2]/t$, where
$e_{\text{site}}$ is the ground-state energy per site and $e_{\text{free}}$ is that
of the free theory and the term~$3\rho G(t)^2$ subtracts the figure-eight
diagram. However, since the subtractions are real,
$\imag\widetilde{\mathcal{E}}=\imag e_{\text{site}}/t$.

\subsection{Strong coupling expansion}
Rescaling $\phi_i\to\rho^{-1/6}\phi_i$, $\pi_i\to\rho^{1/6}\pi_i$ and setting
$\lambda:=\rho^{-2/3}$, we have $H=\lambda^{-1/2}(H_0+V)$ with
\begin{equation}
   H_0=\sum_i\left(\frac{1}{2}\pi_i^2+\phi_i^4\right),\qquad
   V=\alpha\sum_i\phi_i^2+\beta_1\sum_i\phi_i\phi_{i+1}
   +\beta_2\sum_i\phi_i\phi_{i+2},
\label{eq:(B3)}
\end{equation}
where $\alpha=(a_{\text{I}}+b_{\text{I}}+r/2)\lambda$, $\beta_1=-a_{\text{I}}\lambda$
and~$\beta_2=-b_{\text{I}}\lambda$. The unperturbed Hamiltonian $H_0$ consists of
decoupled anharmonic oscillators, and the perturbative expansion in~$V$ is an
expansion in~$\lambda$, i.e.\ a strong coupling expansion.

We organize this expansion by the linked-cluster expansion~\cite{Marland:1981,Hamer:1984,Singh:1988,Oitmaa:2006}. For a finite nonempty set of sites~$c$,
called a cluster, let $E(c)$~be the ground-state energy of~$H_0+V$ in which
only the on-site terms on~$c$ and the bonds both of whose endpoints lie in~$c$
are kept. The cumulant~$W(c)$ is defined recursively by
\begin{equation}
   E(c)=\sum_{\emptyset\neq c'\subseteq c}W(c')
   \quad\Longleftrightarrow\quad
   W(c)=E(c)-\sum_{\emptyset\neq c'\subsetneq c}W(c'),
\label{eq:(B4)}
\end{equation}
starting from single sites, $W(\{i\})=E(\{i\})$. The linked-cluster theorem
states that $W(c)=0$ unless $c$ is connected by the bonds in~$V$. Let $W_\ell$
be the sum of~$W(c)$ over the connected clusters of extent~$\ell$, i.e.\
$c\subseteq\{1,\ldots,\ell\}$ containing both sites $1$ and~$\ell$; because of
the next-nearest-neighbor term in~$V$, $\phi_i\phi_{i+2}$, these are not
necessarily contiguous. Since a cluster of extent~$\ell$ can be placed in a
segment of~$L$ sites in $L-\ell+1$ ways, the ground-state energy of the
segment, $E_L:=E(\{1,\ldots,L\})$, is
\begin{equation}
   E_L=\sum_{\ell=1}^L(L-\ell+1)W_\ell
   \quad\Longleftrightarrow\quad
   W_L=E_L-2E_{L-1}+E_{L-2},
\label{eq:(B5)}
\end{equation}
where $E_0=E_{-1}=0$, and, since the sum over~$\ell$ has an upper bound in a
fixed order of the expansion in~$V$, the ground-state energy per site in the
\emph{infinite} chain is given
by~$\lambda^{1/2}e_{\text{site}}=\lim_{L\to\infty}E_L/L=\sum_{L\geq1}W_L$. Moreover,
since each link between neighboring sites inside a connected cluster must be
crossed by an even number of hopping operators,\footnote{%
Since $H_0$ and its ground-state are invariant under~$\phi_i\to-\phi_i$ at each
site, a nonvanishing contribution must contain an even number of~$\phi_i$ at
every site. Now cut the cluster at a link and count the total number
of~$\phi$'s on the sites to the left of the cut. This number is even. A term
of~$V$ lying entirely on the left contributes two $\phi$'s to it, whereas a
hopping operator crossing the cut contributes only one. Hence the number of
hopping operators crossing the cut is even, and, since the cluster is
connected, it is at least two.} and $\phi_i\phi_{i+2}$ crosses two links, $W_L$
is of order~$\beta_1^{q_1}\beta_2^{q_2}$ with~$q_1+2q_2\geq2(L-1)$.

We compute $E_L$ by the Rayleigh--Schr\"odinger perturbation theory,
$E_L=\sum_{k\geq0}E^{(k)}$, through the recursion
\begin{align}
   E^{(k)}&=\langle\psi^{(0)}|V|\psi^{(k-1)}\rangle,&&
\notag\\
   |\psi^{(k)}\rangle&=R\left[V|\psi^{(k-1)}\rangle
   -\sum_{j=1}^{k-1}E^{(j)}|\psi^{(k-j)}\rangle\right],&
   R&:=\frac{1-|\psi^{(0)}\rangle\langle\psi^{(0)}|}{E^{(0)}-H_0},
\label{eq:(B6)}
\end{align}
where $|\psi^{(0)}\rangle=\prod_i|0\rangle_i$ and~$E^{(0)}=L\varepsilon_0$;
$|n\rangle$ and $\varepsilon_n$ are the $n$th eigenstate and eigenvalue of the
anharmonic oscillator at a site, $h:=\pi^2/2+\phi^4$, which we obtain
numerically. In the basis $\prod_i|n_i\rangle_i$, $H_0=\sum_i\varepsilon_{n_i}$
is diagonal, so that $R$ requires no matrix inversion. The local Hilbert space
of each site is truncated to the lowest $m_L$ eigenstates of~$h$.

We keep the three couplings independent,
\begin{equation}
   \lambda^{1/2}e_{\text{site}}
   =\sum_{p,q_1,q_2}b_{p,q_1,q_2}\,
   \alpha^{p}\beta_1^{q_1}\beta_2^{q_2}
   =\sum_{N\geq0}A_N(r)\lambda^N,
\label{eq:(B7)}
\end{equation}
so that $A_N(r)$ is a polynomial in $r$. We computed $A_N(r)$ to~$N=25$,
keeping $q_2\leq4$, for which $L\leq15$ is required; the results with
$q_2\leq3$, \dots, $6$ agree within $0.1\%$, while a truncation at linear
order in~$\beta_2$ is not sufficient. The truncation of the local Hilbert space
is $m_L=26$, $26$, $22$, $16$, $14$, $11$, $8$, $6$, $5$, $4$, $4$, $3$, $3$,
$3$, $3$ for $L=1$, \dots, $15$, respectively.

\subsection{Resummation}
For~$\rho\to0$, $\lambda^{1/2}e_{\text{site}}\to\lambda^{1/2}e_{\text{free}}$. To
reproduce this weak coupling behavior, we construct the $[(M+1)/M]$ Pad\'e
approximant of~$[\lambda^{1/2}e_{\text{site}}]^2$ and continue it along
$\rho\to|\rho|e^{-i\theta}$, $0\leq\theta\leq\pi$, i.e.\
$\lambda\to|\lambda|e^{2\pi i/3}$~\cite{Nishiyama:2001}. With~$N=25$ we use
$M=10$, $11$ and~$12$, and take the average of the highest two as the value.
A lattice point $(\widetilde{g},t)$ is accepted when (i)~the three approximants
agree within~$12\%$ and are all negative, (ii)~the value agrees with that from
the series truncated at~$N-2=23$ within~$2\%$, and (iii)~all coarser lattices
at the same $\widetilde{g}$ are accepted. Without condition~(ii), the points
near the edge of the accepted region are biased downward. The error of each
point is the combination of the half spread in~(i), the difference in~(ii), and
the error associated with the truncation~$m_L$ (estimated from the change under
$m_9=5\to4$). For~$5\leq\widetilde{g}\leq15$ and~$0.16\leq t\leq2.56$, the
accepted region reaches $t\simeq0.2$ for~$\widetilde{g}\gtrsim12$ and
$t\simeq0.5$ for~$\widetilde{g}=7$.

\subsection{Bounce calculus on the lattice}
The lattice data in~Fig.~\ref{fig:B1}(a) change by a factor of two over the
accessible range of~$t$, and the change depends strongly on~$\widetilde{g}$.
This is because the bounce action itself depends on the lattice spacing and
enters the exponent as~$S/\widetilde{g}$. In order to separate this lattice
artifact, we compute the semiclassical approximation on the same lattice, with
space discrete and the Euclidean time~$\tau$ continuous, as appropriate for the
Hamiltonian~\eqref{eq:(B1)}, and extrapolate only the ratio of the data to it.

Setting $h:=\sqrt{t}$ and~$\phi_i(\tau):=(12/\widetilde{g})^{1/2}u_i(\tau)$, and
rescaling~$\tau\to\tau/h$, we find that the Euclidean action
becomes~$S=(12/\widetilde{g})s[u]$ with
\begin{equation}
   s[u]=\int d\tau\,
   h\sum_i\left[
   \frac{1}{2}(\partial_\tau u_i)^2
   +\frac{1}{2}u_i(-\Delta_{\text{lat}}u)_i
   +\frac{1}{2}u_i^2-\frac{1}{2}u_i^4
   \right],
\label{eq:(B8)}
\end{equation}
where $-\Delta_{\text{lat}}$ is the lattice Laplacian of~Eq.~\eqref{eq:(B1)}
divided by~$h^2$, with the eigenvalues~$\omega_k^2/h^2$.\footnote{%
Note that $h$ is effectively the lattice spacing in the present problem.} The
bounce is the solution of~$-\partial_\tau^2u-\Delta_{\text{lat}}u+u-2u^3=0$
decaying at infinity. In the continuum,
$s[u_{\text{bounce}}]=s_0:=S_0/12=2.92522$. On the lattice, there are two
symmetric solutions, the bounce centered on a site, which has the lowest
action~$s_{\text{lat}}(t)$, and the one centered on the midpoint of a link, which
has a higher action and one more negative mode. The latter is the top of the
barrier that the bounce has to cross when it is moved by one lattice spacing;
the position~$X$ of the bounce is thus not a zero mode but moves in a periodic
Peierls--Nabarro potential~\cite{Peyrard:1984kdi,Speight:2000ed} (see
also~Ref.~\cite{GarciaPerez:1993lic}), whose height,
$\Delta s_{\text{PN}}=s_{\text{link}}-s_{\text{site}}$, decreases as~$e^{-5.6/h}$
for~$h\to0$. Correspondingly, the fluctuation operator around the site-centered
bounce has, besides the negative mode and the zero mode of the
$\tau$~translation, a soft mode of the $x$~translation with the eigenvalue
$\lambda_x\simeq2\pi^2\Delta s_{\text{PN}}/(h^2\sigma_x)$,
where~$\sigma_x=\int d\tau\,h\sum_i(\partial_xu_i)^2$.\footnote{%
For the bounce centered at~$x=X$, we model the lattice action by the periodic
function $s(X)-s(0)=\Delta s_{\text{PN}}\sin^2[(\pi/h)X]$ (this is valid for
small~$t$). Since a shift of the bounce by~$\delta X$ changes $u$
by~$-\delta X\partial_xu$, the second derivative~$\partial_X^2s(X)$ at~$X=0$,
$2\pi^2\Delta s_{\text{PN}}/h^2$, is identified
with~$\int d\tau\,h\sum_i\partial_xu_i\lambda_x\partial_xu_i=\lambda_x\sigma_x$.}
This mode cannot be treated as a Gaussian fluctuation when
$12\Delta s_{\text{PN}}/\widetilde{g}\lesssim1$, and it cannot be treated as an
exact zero mode either, because the weight $e^{-S}$ varies along the path.

We therefore treat the two translations as collective coordinates by the
Faddeev--Popov method with linear gauge conditions. We introduce the gauge
vectors
\begin{equation}
   e_X=(x-X)w,\qquad f=\tau w,\qquad
   w:=\exp\left[-\frac{(x-X)^2+\tau^2}{w_0^2}\right],
\label{eq:(B9)}
\end{equation}
where $x=hi$ ($i$ is the lattice site), and define $u_X$ as the stationary
point of~$s[u]$ under the constraints,
$\langle e_X,u\rangle=\langle f,u\rangle=0$, where the inner product is defined
by~~$\langle a,b\rangle:=\int d\tau\,h\sum_ia_ib_i$. For~$X=0$ and~$h/2$, $u_X$
coincides with the bounces centered on a site and on a link, respectively, and
for~$0<X<h/2$ it interpolates between them along the valley of the action.
Inserting the Faddeev--Popov identity for these conditions into the path
integral and integrating the fluctuations orthogonal to~$e_X$ and~$f$ in the
Gaussian approximation, we obtain the imaginary part per unit length and time
as
\begin{equation}
   \left[\imag\widetilde{\mathcal{E}}\right]_{\text{sc}}
   =-\frac{12}{2\pi\widetilde{g}}\frac{1}{h}\int_0^hdX\,
   J_x(X)J_\tau(X)\left|\det\nolimits_{\perp,{\text{ren}}}(X)\right|^{-1/2}
   e^{-12s[u_X]/\widetilde{g}},
\label{eq:(B10)}
\end{equation}
where
\begin{align}
   J_x&:=\frac{|\langle e_X,\partial_Xu_X\rangle|}
   {\sqrt{\langle e_X,e_X\rangle}},\qquad
   J_\tau:=\frac{|\langle f,\partial_\tau u_X\rangle|}
   {\sqrt{\langle f,f\rangle}},
\notag\\
   \ln\det\nolimits_{\perp,{\text{ren}}}
   &=\ln\frac{\det(K+W)\det(E^T(K+W)^{-1}E)}{\det K\det(E^TE)}
   -\Tr(K^{-1}W).
\label{eq:(B11)}
\end{align}

Here, $K:=-\partial_\tau^2-\Delta_{\text{lat}}+1$, $W:=-6u_X^2$
and~$E:=(e_X,f)$ denotes the rectangular matrix whose two columns are the gauge
vectors, so that $E^TE$ and $E^T(K+W)^{-1}E$ are $2\times2$ matrices with the
elements $\langle a,b\rangle$ and $\langle a,(K+W)^{-1}b\rangle$
($a$, $b\in\{e_X,f\}$), respectively. $J_x$ and~$J_\tau$ are the Faddeev--Popov
Jacobians, and $\det\nolimits_\perp$ is the fluctuation determinant restricted
to the subspace orthogonal to~$e_X$ and~$f$, written as the determinant of a
bordered matrix. The subtraction~$-\Tr(K^{-1}W)$ implements the normal
ordering~\eqref{eq:(B2)}, and the factor $1/h\int_0^hdX$ is the average over
the unit cell. The factor $12/(2\pi\widetilde{g})$ together with $J_xJ_\tau$
replaces the factor $(S_0/2\pi\widetilde{g})^{D/2}$ of Eq.~\eqref{eq:(1.2)}.
Note that Eq.~\eqref{eq:(B10)} contains no free parameter.

For the $\tau$ translation, which is an exact symmetry, Eq.~\eqref{eq:(B10)}
does not depend on the choice of~$f$. For the $x$~translation, the independence
holds only to the extent that the fluctuations orthogonal to the path~$u_X$ can
be integrated in the Gaussian approximation uniformly along the path, i.e.\ as
long as~$\lambda_x$ is small compared with the other fluctuation eigenvalues;
a different gauge vector selects a different path, and the difference is
compensated only in this case. For~$t\to0$, $u_X$ becomes the translated
continuum bounce and Eq.~\eqref{eq:(B10)} reduces to Eq.~\eqref{eq:(1.2)}
with~$C_{2,1}=0.43245$ of~Ref.~\cite{Malatesta:2017}; numerically, the ratios
of Eq.~\eqref{eq:(B10)} to Eq.~\eqref{eq:(1.2)} are~$e^{0.006}$, $e^{0.016}$
and~$e^{0.035}$ at~$t=0.04$, $0.09$ and~$0.16$.


\subsection{Size of lattice effects}
With the improved action, $s_{\text{lat}}-s_0\simeq-0.2t^2$ for small~$t$,
compared with~$\simeq-0.29\,t$ for the standard action ($b_{\text{I}}=0$). The
lattice spacings accessible to the strong coupling expansion are, however,
not small in this sense. The classical action deficit and the Peierls--Nabarro
barrier are
\begin{equation}
   \text{$12(s_0-s_{\text{lat}})=0.44$, $1.88$, $3.59$,
   $12\Delta s_{\text{PN}}=0.26$, $1.83$, $4.16$,
   at $t=0.36$, $0.64$, $1$,}
\label{eq:(B14)}
\end{equation}
both of which enter the exponent divided by~$\widetilde{g}$. The former
enhances the imaginary part and the latter suppresses it; the one-loop factor
in~Eq.~\eqref{eq:(B10)} differs from that of the continuum
by~$3$--$7\%$ for~$0.16\leq t\leq0.36$. At the same time, the soft eigenvalue
grows with~$t$: $\lambda_x\simeq0.09$, $0.20$, $0.38$, $0.57$, and $1.0$
at~$t=0.25$, $0.30$, $0.36$, $0.42$, and~$0.64$, and becomes comparable to the
mass gap. Accordingly, Eq.~\eqref{eq:(B10)} starts to depend on the width~$w_0$
of the gauge vectors~\eqref{eq:(B9)}: when $w_0$ is varied from~$1$ to~$3$,
$\ln\imag\widetilde{\mathcal{E}}_{\text{sc}}$ changes by less than~$0.002$
for~$t\leq0.25$, by~$0.013$ at~$t=0.30$ and $0.03$ at~$t=0.36$, but by~$0.33$
at~$t=0.42$ and by about~$0.7$ for~$t\geq0.64$. In the latter region, the
barrier~$12\Delta s_{\text{PN}}/\widetilde{g}$ is of order unity, so that neither
the collective coordinate (which assumes a soft mode) nor the Gaussian integral
around a pinned bounce (which assumes a high barrier) is controlled. We thus
use the lattice semiclassics only for~$t\leq0.36$.

\subsection{Continuum limit}
\begin{figure}[htbp]
\centering
\includegraphics[width=\textwidth]{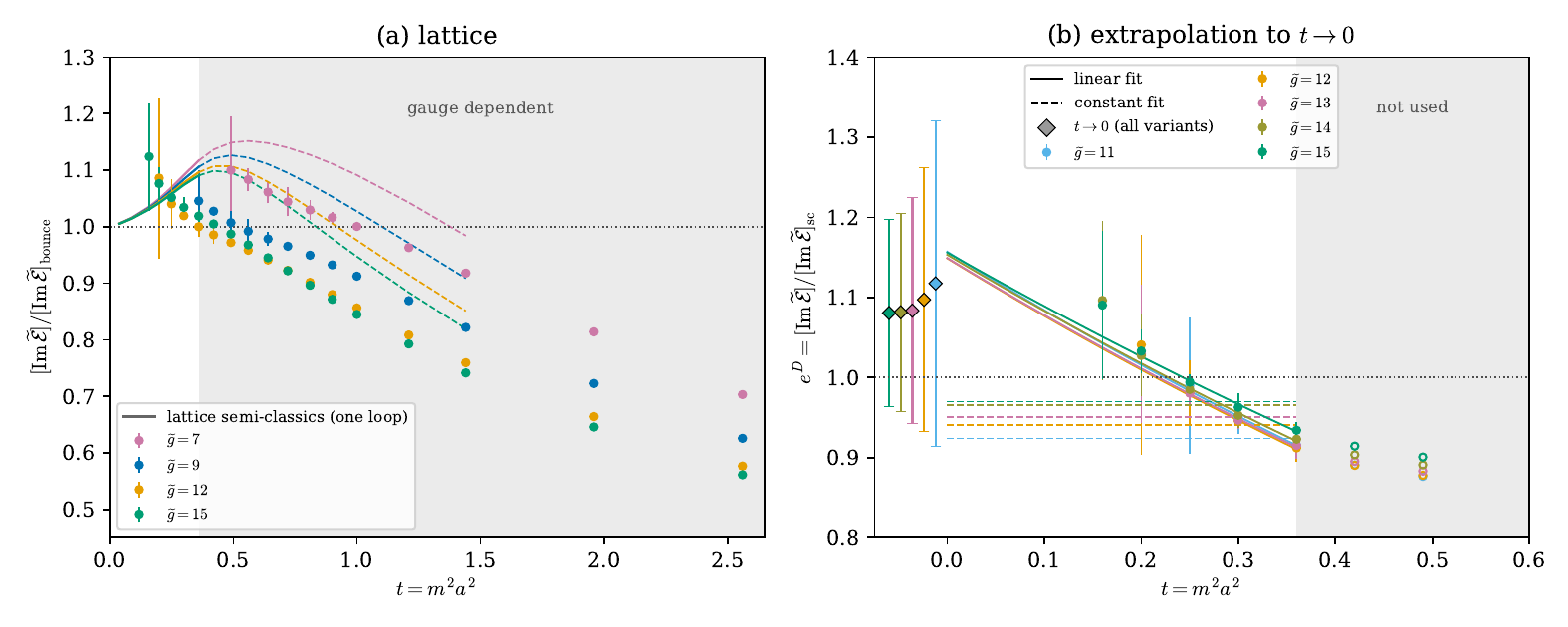}
\caption{(a) The ratio of the imaginary part on the lattice, obtained by the
Pad\'e resummation of the strong coupling series to~$N=25$, to the leading
bounce result~\eqref{eq:(1.2)}, as a function of~$t=m^2a^2$ (points), and the
one-loop semiclassical approximation on the same lattice (lines). In the
shaded region, $t>0.36$, the latter depends on the gauge condition for the
$x$~translation and is shown only for reference (dashed). (b)~The
ratio~$e^D$~\eqref{eq:(B13)}, for~$11\leq\widetilde{g}\leq15$ in the
window~$t\leq0.36$ and its extrapolation to~$t=0$ by a constant (dashed lines)
and by a linear function (solid lines), is shown for the gauge width~$w_0=1$;
the open symbols~($t>0.36$) are not used. The diamonds, slightly displaced
to~$t<0$ for visibility, are the continuum values~\eqref{eq:(B14)}, whose
errors cover all the extrapolation variants.}
\label{fig:B1}
\end{figure}

We define the remainder
\begin{equation}
   e^{D(\widetilde{g},t)}
   =\frac{[\imag\widetilde{\mathcal{E}}]}
   {[\imag\widetilde{\mathcal{E}}]_{\text{sc}}},
\label{eq:(B13)}
\end{equation}
where the numerator is the resummed lattice data. Since Eq.~\eqref{eq:(B10)}
reduces to~\eqref{eq:(1.2)} for~$t\to0$, $e^{D(\widetilde{g},0)}$ is the ratio
$[\imag\widetilde{\mathcal{E}}]/[\imag\widetilde{\mathcal{E}}]_{\text{bounce}}$ in
the continuum, i.e.\ the correction beyond one loop; for~$t>0$ it also
contains lattice artifacts beyond one loop. Although the data themselves vary
by a factor of two, $e^D$~stays close to unity in the window~$t\leq0.36$ where
Eq.~\eqref{eq:(B10)} is reliable, and decreases from about~$1.09$ at~$t=0.16$
to about~$0.92$ at~$t=0.36$ (Fig.~\ref{fig:B1}(b)). Only the stronger
couplings, $\widetilde{g}\gtrsim9$, have accepted points in this window, and we
quote the continuum value for~$11\leq\widetilde{g}\leq15$, which have three to
five points each. The functional form of the remaining $t$ dependence is not
known; the linear function of~$t$ fits the data well
($\chi^2/{\text{dof}}\leq0.14$), while a constant does not
($\chi^2/{\text{dof}}=1.5$--$6.6$), but the window is short and its lower end,
$t=0.16$--$0.2$, has the largest errors. We therefore extrapolate $D$ to~$t=0$
both by a constant and by a linear function of~$t$, for the four widths
$w_0=1$, $1.5$, $2$ and $3$, and take the midpoint of the range spanned by
these variants, including their fit errors, as the value and the half width as
the error. This gives
\begin{equation}
   \text{
   $\frac{[\imag\widetilde{\mathcal{E}}]}
   {[{\imag\widetilde{\mathcal{E}}}]_{\text{bounce}}}
   =1.12(20)$, $1.10(17)$, $1.08(14)$, $1.08(12)$, $1.08(12)$
   for $\widetilde{g}=11$, \dots, $15$.}
\label{eq:(B14)}
\end{equation}
The errors are dominated by the choice of the extrapolation: the constant fits
give $0.92$--$0.98$ and the linear fits $1.08$--$1.16$. This is consistent with
the main text, $0.95$--$0.97$ for~$P=8$. The present accuracy, however, does
not allow a quantitative test; such a test at the level of a few percent would
require the strong coupling series to reach~$t\lesssim0.1$, i.e.\ considerably
higher orders than~$N=25$.

Finally, we examined the effect of the Symanzik improvement beyond the tree
level. At one loop, the self-energy is the tadpole, which is independent of the
momentum and is removed exactly by the normal ordering~\eqref{eq:(B2)}, so that
$b_{\text{I}}$ is not corrected. The remaining $O(a^2)$~artifact at this order is
that of the four-point function, i.e.\ of the one-loop bubble at zero external
momentum,
\begin{equation}
   B_{\text{lat}}(t)
   =\int_{-\pi}^\pi\frac{dk}{2\pi}\,\frac{1}{4(\omega_k^2+t)^{3/2}},
   \qquad
   c:=\lim_{t\to0}\left[B_{\text{lat}}(t)-\frac{1}{4\pi t}\right]=0.00166,
\label{eq:(B15)}
\end{equation}
where $1/(4\pi t)$ is the continuum value in lattice units; it is removed by
the replacement $g\to g\,[1-(3/2)cga^2]$ of the bare coupling, i.e.\
$\rho\to\rho(1-(3/2)c\,\widetilde{g}t)$ in~Eqs.~\eqref{eq:(B1)}
and~\eqref{eq:(B2)}. This changes the lattice data by
$\delta\ln[\imag\widetilde{\mathcal{E}}]\simeq-(3/2)ct(S_0-\widetilde{g})$,
which is at most~$-0.02$ in the window~$t\leq0.36$. Since the lattice
semiclassics~\eqref{eq:(B10)} evaluated with the same bare coupling changes by
the same amount to this order, $D$~changes by less than~$0.003$ and the
values~\eqref{eq:(B14)} by about~$0.1\%$; even if only the data are improved,
the values~\eqref{eq:(B14)} decrease by less than $1\%$.




%



\let\doi\relax









\end{document}